\documentclass[aps,pra,reprint,superscriptaddress,longbibliography]{revtex4-1}

\usepackage{epstopdf}
\usepackage{amsmath,amssymb,graphicx,bm}
\usepackage{graphicx}
\usepackage{color}
\usepackage{dcolumn}
\usepackage{bm}

\begin{document}


\title{Interplay between diffraction and the Pancharatnam-Berry phase in inhomogeneously twisted anisotropic media}



\author{Chandroth P. Jisha}
\email[]{cpjisha@gmail.com}
\affiliation{Centro de F\'{\i}sica do Porto, Faculdade de Ci\^encias, Universidade do Porto, PT-4169-007 Porto, Portugal}

\author{Alessandro Alberucci}
\email[]{alessandro.alberucci@tut.fi}
\affiliation{Optics Laboratory, Tampere University of Technology, FI-33101 Tampere, Finland}
\affiliation{N\textsl{oo}EL - Nonlinear Optics and OptoElectronics Lab, University ``Roma Tre'', IT-00146 Rome, Italy}

\author{Lorenzo Marrucci}
\affiliation{Dipartimento di Fisica, Universit\`a di Napoli Federico II, IT-80100 Naples, Italy}
\affiliation{CNR-ISASI, Institute of Applied Science and Intelligent Systems, IT-80078 Pozzuoli (NA), Italy }

\author{Gaetano Assanto}
\affiliation{Optics Laboratory, Tampere University of Technology, FI-33101 Tampere, Finland}
\affiliation{N\textsl{oo}EL - Nonlinear Optics and OptoElectronics Lab, University ``Roma Tre'', IT-00146 Rome, Italy}
\affiliation{CNR-ISC, Institute for Complex Systems, IT-00185 Rome, Italy}

\date{\today}

\begin{abstract}
We discuss the propagation of an electromagnetic field in an inhomogeneously anisotropic material where the optic axis is rotated in the transverse plane but is invariant along the propagation direction. In such a configuration, the evolution of an electromagnetic wavepacket is governed by the Pancharatnam-Berry phase (PBP), responsible for the appearance of an effective photonic potential. In a recent paper [A. Alberucci et al., ``Electromagnetic confinement via spin-orbit interaction in anisotropic dielectrics'',  ACS Photonics \textbf{3},  2249 (2016)] we demonstrated that the effective potential supports transverse confinement. Here we find the profile of the quasi-modes and show that the photonic potential arises from the Kapitza effect of light. The theoretical results are confirmed by numerical simulations, accounting for the medium birefringence. Finally, we analyze in detail a configuration able to support non-leaky guided modes.
\end{abstract}


\maketitle

\section{Introduction}

In the last few years, a great deal of attention has been devoted to investigating light propagation in inhomogeneously anisotropic materials, featuring a rotation of the optic axis but no variations in the local refractive indices. In such configuration, a phase distribution proportional to the local rotation angle of the optic axis can be superposed to the beam wavefront, leading to the so-called planar photonics \cite{Kildishev:2013,Yu:2014,Lin:2014,Kobashi:2016}. Such space-dependent phase delay is due to a gradient in the Pancharatnam-Berry phase (PBP) \cite{Bomzon:2002,Marrucci:2006_1}, the latter being associated with a change in the beam polarization. For a closed path (same initial and final state) the PBP is provided by the solid angle subtended by the closed circuit on the Poincar\'e sphere \cite{Berry:1987}. \\
Historically, the idea of modulating the phase of an electromagnetic wave via the PBP is due to Bomzon and coworkers, who used a metallic sub-wavelength grating (nowadays it would be called a two-dimensional metamaterial) \cite{Bomzon:2001,Hasman:2003}. This concept was then applied by Marrucci and collaborators to nematic liquid crystals (NLCs), natural materials where the pointwise orientation of the optic axis can be engineered by tailoring the boundary conditions via optical alignment techniques \cite{Marrucci:2006,Slussarenko:2011,Kim:2015}. This full control over the local NLC director paved the way to new functionalities, for example the conversion of spin angular momentum (polarization degree of freedom) to orbital angular momentum (spatial degree of freedom) \cite{Bauer:2015,Naidoo:2016} and the realization of polarization gratings in both amplitude \cite{Todorov:1992,Gori:1999} and phase \cite{Provenzano:2007,Tabiryan:2011,Ruiz:2013}. 
Later on, the interest on PBP was boosted thanks to the introduction of metasurfaces and the demonstration of polarization-dependent optical devices \cite{Lin:2014,Litchinitser:2016}, such as lenses \cite{Roux:2006,Khorasaninejad:2016}, holograms \cite{Zheng:2015}, vortex plates in polymerizable NLCs \cite{Nersisyan:2009}, deflectors based on the spin-Hall effect \cite{Li:2013,Ling:2015}. Recently, the PBP has been used to tailor the beam wavefront upon reflection from a layer of chiral liquid crystals \cite{Rafayelyan:2016,Kobashi:2016,Kobashi:2016_1,Barboza:2016}.\\
In this Paper we investigate the propagation of finite-size electromagnetic wavepackets in inhomogeneously anisotropic materials, subject to a pointwise rotation of the principal axes (Fig.~\ref{fig:geometry_3D}) but invariant along the propagation direction $z$. Previously, using geometric optics other authors analyzed the dependence of the beam trajectory on the photon for odd distribution of the rotation angle, the so called optical Magnus effect or spin-Hall effect \cite{Bliokh:2007,Bliokh:2008}. For even rotations, the interplay between diffraction and PBP was discussed for wide beams as superpositions of plane waves in Refs.~\cite{Calvo:2007,Karimi:2009}, whereas the approach we undertake hereby holds valid until longitudinal field components become relevant, i.e., for beams of size comparable to or narrower than the wavelength \cite{Lax:1975}. In a recent Letter we showed that light propagates under the influence of an effective photonic potential, leading to leaky guided modes for bell-shaped distributions of the optic axis rotation  \cite{Alberucci:2016} (Fig.~\ref{fig:geometry_3D}). In this Article we provide a complete and detailed description of the theory -including a quantitative comparison with numerical simulations- first published in \cite{Alberucci:2016}. The paper is structured as follows. In Section~\ref{sec:thin_cells} we briefly recall the basics of optical propagation in twisted anisotropic materials in the absence of diffraction. In Section~\ref{sec:long_sample}  we generalize to the three-dimensional case the paraxial equations governing light propagation found in Ref.~\cite{Alberucci:2016} to the case of long (with respect to the Rayleigh length) samples. We also sketch out the comparison with Pauli equation for quantum particles. In Section~\ref{sec:quasi_modes} we first discuss the physical reason behind the absence of longitudinally-invariant modes, and then we find the equations governing the quasi-modes, including the higher-order components of the localized waves. In Section~\ref{sec:origin} we provide an extended physical explanation about the origin of the photonic potential as a Kapitza effect, i.e. a periodic modulation of the PBP versus propagation providing a $z$-independent effective photonic potential \cite{Kapitza:1951,Alberucci:2013}, by using the plane-wave solution shown in Section~\ref{sec:thin_cells}. In Section~\ref{sec:potentials} first the relationship between the symmetry of the twisting angle and light behavior is recalled, and then the profile of the CW (continuous wave) component of the quasi-mode is computed for the first time. In Section~\ref{sec:simulations} we check the validity of the theoretical results using two different types of numerical simulations, based on an in-house BPM (Beam Propagation Method) code written in the rotated reference system and on an open-access FDTD (Finite-Difference Time Domain) code. Through numerical simulations we check the validity of effective photonic potential and -in the guiding case- the accuracy of the associated quasi-modes found in Section~\ref{sec:potentials}. In Section~\ref{sec:nonleaky} we report the design of a non-leaky (V-shaped) waveguide in a twisted structure and test its properties via FDTD simulations. Finally, in Section~\ref{sec:conclusions} we summarize the contents of the Article, pinpointing the novelties with respect to Ref.~\cite{Alberucci:2016} and illustrating future developments, both on the theoretical and on the experimental/technological sides.

\section{Pancharatnam-Berry phase in a thin sample}
\label{sec:thin_cells}
\begin{figure}
\includegraphics[width=0.5\textwidth]{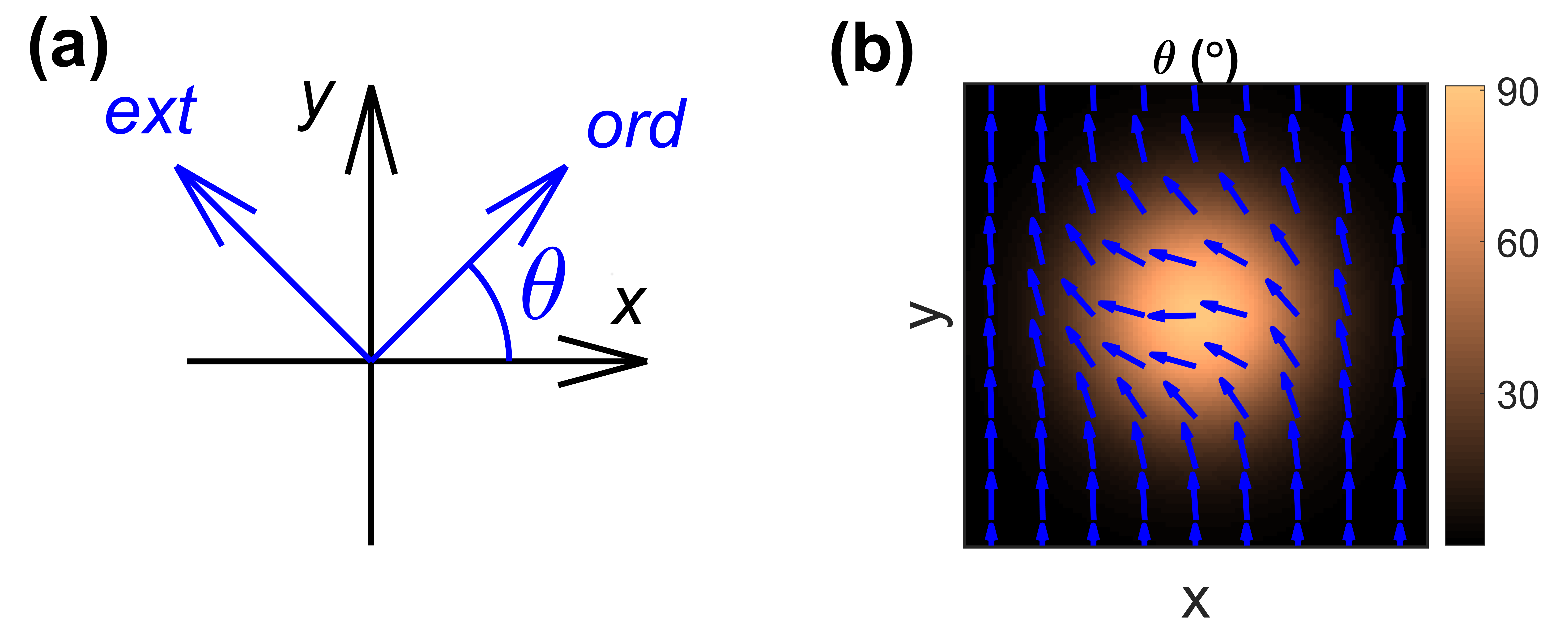}
\caption{\label{fig:geometry_3D} (a) Definition of the rotation angle $\theta$. (b) Gaussian distribution of the rotation angle $\theta$ on the plane $xy$ when maximum rotation is $90^\circ$: the arrows correspond to the local optic axis (i.e., the extraordinary axis), superposed to the spatial distribution of $\theta$ represented as a color map. When $\delta=\pi$, the distribution plotted in (b) corresponds to a polarization-dependent lens \cite{Roux:2006,Lin:2014}.}
\end{figure}
We consider a non-magnetic anisotropic uniaxial crystal \footnote{\textit{Mutatis mutandis}, all the following results remain valid in the more general case of a biaxial crystal.}. In the reference system of the principal axes (subscript $D$), the relative permittivity is
\begin{equation}
  \bm{\epsilon}_D =\left( \begin{array}{ccc} 
	\epsilon_\bot & 0 & 0 \\
	0 & \epsilon_\| & 0 \\
	0 & 0 &\epsilon_\bot \\
	\end{array}\right),
\end{equation} 
where $\epsilon_\bot$ and $\epsilon_\|$ are the dielectric constants for fields oscillating normal and parallel to the optic axis $\hat{n}$, respectively. The optic axis $\hat{n}$ is normal to the wavevector $\bm{k}$ and at an angle $\theta(x,y)$ with respect to the axis $y$ [Fig.~\ref{fig:geometry_3D}(a)]. Noteworthy, $\epsilon_\bot$ and $\epsilon_\|$ are spatially uniform: this condition, together with the previous assumption $\hat{n}\cdot \bm{k}=0$, rules out any refractive index changes in the transverse plane \cite{Slussarenko:2016}. We then introduce the ordinary and extraordinary refractive indices as $n_\bot=\sqrt{\epsilon_\bot}$ and $n_\|=\sqrt{\epsilon_\|}$, respectively. The local rotation of the optic axis (i.e., of the principal axes) around $z$ can be expressed as 
\begin{equation}  \label{eq:rotation}
  \bm{R}(\theta) = \left ( \begin{array} {ccc}
  \cos\theta & \sin\theta & 0 \\
  -\sin\theta & \cos\theta & 0 \\ 
	 0 & 0 & 1\\
 \end{array} \right).
\end{equation} 
The matrix \eqref{eq:rotation} operates on a vector in the laboratory framework $xyz$ and yields its coordinates in the Cartesian system aligned with the local principal axes. The relative permittivity $\bm{\epsilon}$ in $xyz$ is then provided by
\begin{equation} \label{eq:epsilon_rotated}
   \bm{\epsilon}(x,y)=\bm{R}[-\theta(x,y)]\cdot \bm{\epsilon}_D \cdot \bm{R}[\theta(x,y)].
\end{equation}
When diffraction in the anisotropic material can be neglected, light propagation can be modeled by using plane waves. In fact, the Jones formalism \cite{Jones:1941} can be employed accounting for the transmission dependance on the transverse coordinate $(x,y)$ through the rotation angle $\theta$. In the plane-wave approximation, light propagation through an anisotropic slab of length $z$ is given by 
\begin{equation}
  \left( \begin{array}{c}  E_o(x,y,z) \\ E_e(x,y,z) 
	\end{array} \right)= \left ( \begin{array} {cc}
  e^{ik_0 n_\bot z} & 0  \\
   0 & e^{ik_0 n_\| z}  \end{array} \right) \cdot 
	\left( \begin{array}{c}  E_o(x,y,0) \\ E_e(x,y,0) 
	\end{array} \right),
\end{equation}
where $E_o$ and $E_e$ are the local ordinary and extraordinary polarization components, respectively [Fig.~\ref{fig:geometry_3D}(a)]. \\
Since ordinary and extraordinary directions vary with $(x,y)$, it is more convenient to refer to the two circularly polarized beams which are eigensolutions of the rotation operator \eqref{eq:rotation}. We thus introduce the circular polarization basis $\hat{L}=(\hat{x}-i\hat{y})/\sqrt{2}$ (LCP, left-circular polarization) and $\hat{R}=(\hat{x}+i\hat{y})/\sqrt{2}$ (RCP, right-circular polarization) \footnote{In this Paper we adopt the the source's point of view.}. The basis transformation from linear to circular  is obtained by the matrix $\bm{P}$ 
\begin{equation}
  \left( \begin{array}{c}  E_L \\ E_R 
	\end{array} \right)= \bm{P}\cdot \left( \begin{array}{c}  E_o \\ E_e 
	\end{array} \right) = \frac{1}{\sqrt{2}} \left ( \begin{array} {cc}
  1 & -i  \\
   1 & i  \end{array} \right) \cdot 
	\left( \begin{array}{c}  E_o \\ E_e 
	\end{array} \right),
\end{equation}
where the subscripts L and R correspond to LCP and RCP, respectively.
Finally, the overall transmission of the anisotropic slab is \cite{Slussarenko:2016}
\begin{align}
   &\left( \begin{array}{c}  E_L(x,y,z) \\ E_R(x,y,z) 
	\end{array} \right) = 
	\nonumber \\
	 &e^{i\bar{n}k_0z} \left ( \begin{array} {cc}
    \cos\left(\frac{\delta}{2}\right) & -i\sin\left(\frac{\delta}{2}\right)e^{2i\theta}  \\
   -i\sin\left(\frac{\delta}{2}\right)e^{-2i\theta} & \cos\left(\frac{\delta}{2}\right)  \end{array} \right) \cdot 
	\left( \begin{array}{c}  E_L(x,y,0) \\ E_R(x,y,0) \label{eq:PW}
	\end{array} \right) ,  
\end{align}
where $\delta(z)=k_0 z\Delta n$ is the retardation between ordinary and extraordinary components, $\Delta n = n_\| - n_\bot$ is the birefringence and $\bar{n}=(n_\| + n_\bot)/2$ is the average index of refraction \cite{Marrucci:2006}. Eq.~\eqref{eq:PW} describes a continuous exchange of power between the two circular polarizations RCP and LCP due to  birefringence. Considering a purely circular polarization at the input $z=0$ by setting, for example, $E_R(x,y,0)=1$ and $E_L(x,y,0)=0$. When $\delta=(2l+1)\pi$, with $l$ an integer, the RCP wave turns into LCP: this change in polarization state is accompanied by a dynamic phase change $\Delta\phi_\mathrm{dyn}=(2l+1)\pi \bar{n}/\Delta n$ and a further phase shift $\Delta\phi_\mathrm{geo}=\pm 2\theta$ purely of  geometric origin, a manifestation of PBP \cite{Marrucci:2006}. Planar photonics elements based on PBP usually work  in the latter regime, i.e.,  with the length of the anisotropic material designed to achieve a $\pi$ phase delay between  ordinary and extraordinary components. 

\section{Pancharatnam-Berry phase in an extended sample}
\label{sec:long_sample}
Neglecting anisotropy in the diffraction operator \cite{Kwasny:2012}, 
Maxwell's equations in the paraxial approximation read
\begin{equation} \label{eq:maxwell_equation}
 \nabla^2 \left ( \begin{array} {c}
  E_x \\
  E_y \\
 \end{array} \right)   + k_0^2 \left ( \begin{array} {cc}
  \epsilon_{xx}(x,y) & \epsilon_{xy}(x,y) \\
  \epsilon_{yx}(x,y) & \epsilon_{yy}(x,y) \\
 \end{array} \right) \left ( \begin{array} {c}
  E_x \\
  E_y \\
 \end{array} \right)=0,
\end{equation}
where the relative permittivity is given by \eqref{eq:epsilon_rotated}. The paraxial approximation in Eq.~\eqref{eq:maxwell_equation} allows one to neglect the longitudinal electric field whenever the beam size exceeds the wavelength \cite{Lax:1975}. \\
We apply the slowly varying envelope approximation through the transformation $E_o=e^{ik_0 n_\bot z}\psi_o$ and $E_e=e^{ik_0 n_\| z}\psi_e$, i.e., factoring out the dynamic phase responsible for polarization rotation versus propagation. For paraxial wavepackets, the second derivatives along $z$ can be neglected and Eq.~\eqref{eq:maxwell_equation} yields \cite{Alberucci:2016}
\begin{widetext}
\begin{align}
    2ik_0 n_\bot \frac{\partial {\psi}_o}{\partial z}  &=
 -   \nabla_t^2 \psi_o +  \left[ \left(\frac{\partial \theta}{\partial x} \right)^2 + \left(\frac{\partial\theta}{\partial y} \right)^2 \right] \psi_o  +  \left( \frac{\partial^2 \theta}{\partial x^2} + \frac{\partial^2 \theta}{\partial y^2}  \right)   \psi_e e^{ik_0\Delta n z} + 2 \left( \frac{\partial\theta}{\partial x} \frac{\partial {\psi_e}}{\partial  x} + \frac{\partial\theta}{\partial y} \frac{\partial {\psi_e}}{\partial  y} \right) e^{ik_0\Delta n z}, \label{eq:SVEA_ord} \\
2ik_0 n_\| \frac{\partial {\psi}_e}{\partial z}  &=
 -   \nabla_t^2{\psi_e} +  \left[ \left(\frac{\partial \theta}{\partial x} \right)^2 + \left(\frac{\partial\theta}{\partial y} \right)^2 \right] \psi_e  - \left( \frac{\partial^2 \theta}{\partial x^2} + \frac{\partial^2 \theta}{\partial y^2}  \right)  \psi_o e^{-ik_0\Delta n z}  - 2 \left( \frac{\partial\theta}{\partial x} \frac{\partial {\psi_o}}{\partial  x} + \frac{\partial\theta}{\partial y} \frac{\partial {\psi_o}}{\partial  y} \right) e^{-ik_0\Delta n z}, \label{eq:SVEA_ext}
	\end{align}
\end{widetext}
where $\nabla^2_t=\partial^2_x+\partial^2_y$. Equations (\ref{eq:SVEA_ord}-\ref{eq:SVEA_ext}) indicate that the waves are not subject to refractive index gradients, as transverse modulation is only due to the pointwise rotation of the principal axes. For small birefringence, i.e., $n_\bot\approx n_\|$, Eqs.~(\ref{eq:SVEA_ord}-\ref{eq:SVEA_ext}) resemble Pauli equation for a  charged massive particle moving in a bidimensional space \cite{Dirac:1999}  
\begin{equation} \label{eq:pauli}
i\hbar\frac{\partial \bm{\psi}}{\partial t}=-\frac{\hbar^2}{2 m}\nabla_{xy}^2 \bm{\psi} + U(x,y)\bm{I} \cdot \bm{\psi} + \bm{H}_{LS}(x,y,t) \cdot \bm{\psi},
\end{equation}
with $\bm{I}$ the identity operator  and $\bm{H}_{LS}$ the Hermitian matrix responsible for spin-orbit coupling \cite{Bliokh:2015} \footnote{From the standard analogy between 2D quantum mechanics and paraxial optics in the monochromatic regime, the propagation coordinate $z$plays the role of time.}; $\bm{\psi}$ is a two-component spinor with elements $\psi_o$ and $\psi_e$, respectively. The term containing $U(x)$ is a scalar potential acting equally on both components; the term with $\bm{H}_{LS}(x,y,t)$, proportional to the Pauli matrix $\bm{S}_2$, can be associated to an equivalent time-dependent magnetic field responsible for spin-rotation (in our case power exchange between extraordinary and ordinary components).

\section{Quasi-modes in (1+1)D}
\label{sec:quasi_modes}
Equations~(\ref{eq:SVEA_ord}-\ref{eq:SVEA_ext}) do not support $z$-invariant modes due to the explicit dependence on the propagation coordinate $z$. This is due to the unavoidable power exchange between  ordinary and  extraordinary components when the optic axis rotates across the transverse plane. Noteworthy, the coupling between the two components does not vanish for any coordinate transformation. To clarify this, let us take a given point $P=(x_P,y_P)$ in the transverse plane with $\theta_P=\theta(P)$ and assume the wavepacket to be, e.g., purely extraordinary. Due to diffraction, wavelets in $P$ will spread towards adjacent points $P+dP$, where $\theta(P+dP)=\theta_P+d\theta$ due to the optic axis rotation. In $P+dP$, light will then be in a superposition of ordinary and extraordinary polarization states, no matter how small is the change in $\theta$. This can also be explained through a direct analogy with the quantum mechanics of spinning particles: according to Eq.~\eqref{eq:pauli}, photons behave like spin 1/2 particles \cite{Ballantine:2016} subject to a magnetic field rotating in the plane $xy$ versus the propagation coordinate $z$. 
Thus, the commutation rule between orthogonal magnetic fields forbids the existence of stationary/energy eigenstates \cite{Dirac:1999}. Non-Abelian propagation of light in inhomogeneously anisotropic materials was discussed earlier in the context of geometric optics \cite{Bliokh:2008}. \\
From what stated above, no eigenmodes of Eqs.~(\ref{eq:SVEA_ord}-\ref{eq:SVEA_ext}) exist, as confirmed by  direct numerical investigations. Nonetheless, the system could support quasi-modes, i.e., periodically-varying solutions of finite lateral extension \cite{Shirley:1965,Brown:1991}. Hereafter, for the sake of simplicity we consider the one-dimensional case with $\partial_y=0$. The periodic quasi-modes of Eqs.~(\ref{eq:SVEA_ord}-\ref{eq:SVEA_ext}) can be calculated with the ansatz
\begin{equation}  \label{eq:ansatz_quasi_modes}
 \psi_j=g_j(x)e^{i \sum_{p=-\infty}^\infty {\int \overline{\beta}_p^{(j)}(x,z)dz}}\ (j=o,e),
\end{equation}
 where $g_j(x)$ are $z$-independent functions, whereas the complex exponentials account for spatial variations - in both phase and amplitude - on longitudinal scales equal or smaller than the beat-length $\lambda/\Delta n$. The periodicity of the quasi-mode is ensured by $\overline{\beta}^{(j)}_p(x,z)=\beta^{(j)}_p(x)\exp{\left(\frac{i2 \pi p \Delta n }{\lambda}z\right)}$ \cite{Shirley:1965}. \\
To calculate the quasi-modes of a  guiding PBP structure, we start from Eqs.~(\ref{eq:SVEA_ord}-\ref{eq:SVEA_ext}) with the ansatz \eqref{eq:ansatz_quasi_modes} and find
\begin{widetext}
\begin{align}
  &-2k_0 n_\bot g_o \sum_p \overline{\beta}_p^{(o)}(x,z)= - \left\{ \frac{d^2 g_o}{dx^2} +2i\frac{d g_o}{dx}\frac{\partial \phi_o}{\partial x} + g_o \left[ i \frac{\partial^2 \phi_o}{\partial x^2} -\left( \frac{\partial \phi_o}{\partial x}\right)^2 \right]  \right\} + \left( \frac{d\theta}{dx}\right)^2 g_o \nonumber \\
	& + \left[ \frac{d^2 \theta}{dx^2} g_e  + 2 \frac{d \theta}{dx} \left(\frac{dg_e}{dx} + i g_e  \frac{\partial \phi_e}{\partial x} \right) \right] e^{i k_0\Delta n z}e^{i(\phi_e-\phi_o)}, \label{eq:ORD}\\
	 &-2k_0 n_\| g_e \sum_p \overline{\beta}_p^{(e)}(x,z)= - \left\{ \frac{d^2 g_e}{dx^2} +2i\frac{d g_e}{dx}\frac{\partial \phi_e}{\partial x} + g_e \left[ i \frac{\partial^2 \phi_e}{\partial x^2} -\left( \frac{\partial \phi_e}{\partial x}\right)^2 \right]  \right\} + \left( \frac{d\theta}{dx}\right)^2 g_e \nonumber \\
	& - \left[ \frac{d^2 \theta}{dx^2} g_o  + 2 \frac{d \theta}{dx} \left(\frac{dg_o}{dx} + i g_o  \frac{\partial \phi_o}{\partial x} \right) \right] e^{-i k_0\Delta n z}e^{i(\phi_o-\phi_e)}, \label{eq:EXT}
\end{align}
\end{widetext}
where we introduced $\phi_j(x,z)=\sum_p\int{\overline{\beta}_p^{(j)}(x,z)dz}$. A quasi-mode defined by \eqref{eq:ansatz_quasi_modes} conserves its profile in propagation provided that $\frac{d\beta_0^{(j)}}{dx}=0$ is satisfied. \\
Equations~(\ref{eq:ORD}-\ref{eq:EXT}) contain periodic terms on both sides, thus each harmonic can be equalized separately. The computation is much easier if we assume that $\phi_o-\phi_e$ is constant versus $z$: after the derivation we will verify \textit{a posteriori} that the latter condition is satisfied to first order. \\
Focusing on the cw components $p=0$, the latter satisfy the eigenvalue problem
\begin{align}
   -2\beta_0^{(o)}k_0 n_\bot & {g}_o =
 -   \frac{d^2 {g_o}}{dx^2} + \nonumber \\ & \left[ \left(\frac{d\theta}{dx} \right)^2 + \frac{1}{k_0^2\left(\Delta n\right)^2} \sum_{p=1}^\infty{\frac{1}{p^2}} \frac{d \beta_p^{(o)}}{d x} \frac{d \beta_{-p}^{(o)}}{dx} \right] g_o, \label{eq:eig_ord} \\
-2\beta_0^{(e)}k_0 n_\| & {g}_e =
 -   \frac{d^2 {g_e}}{dx^2} + \nonumber \\ & \left[ \left(\frac{d\theta}{dx} \right)^2 + \frac{1}{k_0^2\left(\Delta n\right)^2} \sum_{p=1}^\infty{\frac{1}{p^2}} \frac{d \beta_p^{(e)}}{dx} \frac{d\beta_{-p}^{(e)}}{dx} \right] g_e. \label{eq:eig_ext}
\end{align}
Thus  Eqs.~(\ref{eq:eig_ord}-\ref{eq:eig_ext}) show that the terms coupling extraordinary and ordinary waves in Eqs.~(\ref{eq:SVEA_ord}-\ref{eq:SVEA_ext}) act on the cw components $g_j(x)\ (j=o,e)$ as additional contributions (summations over $p$) to the photonic potential. Such higher-order contributions, however, from a physical point of view are expected to be relevant only when the beat-length $L_B=\lambda/\Delta n$ exceeds the Rayleigh distance $L_R=\pi\overline{n}w^2_m/\lambda$, with $w_m$ the transverse extent of the localized mode. \\
 When higher-order contributions can be neglected (i.e., $L_B/L_R\ll 1$), Eqs.~(\ref{eq:eig_ord}-\ref{eq:eig_ext}) describe a wavepacket evolving under the action of a photonic potential \footnote{Given the standard paraxial equation $2ik_0\overline{n}\frac{\partial A}{\partial z}+ \frac{\partial^2 A}{\partial x^2}+k_0^2 \Delta n^2 A=0$, the photonic potential is defined as $V
=-k_0 \frac{\Delta n^2}{2\overline{n}}$. } 
\begin{equation} \label{eq:photonic_potential}
V(x)= \frac{1}{2n_j k_0}\left(\frac{d\theta}{dx} \right)^2.
\end{equation} 
We need now to evaluate the degree of accuracy in using the potential \eqref{eq:photonic_potential} into Eqs.~(\ref{eq:eig_ord}-\ref{eq:eig_ext}). Equations~(\ref{eq:ORD}-\ref{eq:EXT})  also determine the profile of the functions $\beta_p^{(j)}(x)$ for $p\neq 0$,  yielding the periodic evolution of the quasi-mode along $z$.  The knowledge of $\beta_p^{(j)}(x)$  allows evaluating the weight of the term $\propto \sum_{p=1}^\infty{\frac{1}{p^2}} \frac{d\beta_p^{(o)}}{dx} \frac{d\beta_{-p}^{(o)}}{dx} $ on the photonic potential. For the sake of simplicity, assuming non-negligible $\beta_p^{(j)}$  only for $\left|p\right|\leq 1$, for $p=1$ we get
\begin{align}
  &-2k_0 n_\bot g_o \beta_1^{(o)}= -\frac{\lambda}{\pi\Delta n} \frac{dg_o}{dx}\frac{d \beta_1^{(o)}}{d x} - \frac{\lambda}{2\pi\Delta n} g_o \frac{d^2 \beta_1^{(o)}}{d x^2} \nonumber \\ &+  g_e \frac{d^2 \theta}{dx^2}  +2 \frac{dg_e}{d x} \frac{d\theta}{dx}, \label{eq:beta_o1} \\
	&-2k_0 n_\| g_e \beta_1^{(e)}= -\frac{\lambda}{\pi\Delta n} \frac{dg_e}{dx}\frac{d \beta_1^{(e)}}{d x} - \frac{\lambda}{2\pi\Delta n} g_e \frac{d^2 \beta_1^{(e)}}{d x^2}. \label{eq:beta_e1}
\end{align}
Similarly, for $p=-1$
\begin{align}
  &-2k_0 n_\bot g_o \beta_{-1}^{(o)}= \frac{\lambda}{\pi\Delta n} \frac{dg_o}{dx}\frac{d \beta_{-1}^{(o)}}{d x} + \frac{\lambda}{2\pi\Delta n} g_o \frac{d^2 \beta_{-1}^{(o)}}{d x^2}, \label{eq:beta_ominus1} \\ 
	&-2k_0 n_\| g_e \beta_{-1}^{(e)}= \frac{\lambda}{\pi\Delta n} \frac{dg_e}{dx}\frac{d\beta_{-1}^{(e)}}{d x} + \frac{\lambda}{2\pi\Delta n} g_e \frac{d^2 \beta_{-1}^{(e)}}{d x^2} \nonumber \\  & -  g_o \frac{d^2 \theta}{dx^2}  - 2 \frac{d g_o}{d x} \frac{d\theta}{dx} . \label{eq:beta_eminus1}  
\end{align}
In the general case the functions $\beta_{p}^{(j)}$ satisfy a second-order ordinary differential equation with coefficients given by the cw component $g_j(x)$. A drastic approximation is at hand for large $\Delta n$, when Eqs.~(\ref{eq:beta_o1}-\ref{eq:beta_eminus1}) provide $\beta_1^{(e)}=\beta_{-1}^{(o)}=0$, $\beta_1^{(o)}=-\left.\left( g_e \frac{d^2 \theta}{dx^2} +2\frac{d\theta}{dx}\frac{dg_e}{dx} \right)\right/ \left(2k_0 n_\bot g_o \right)$ and $\beta_{-1}^{(e)}=\left.\left( g_o \frac{d^2 \theta}{dx^2} +2\frac{d\theta}{dx}\frac{dg_o}{dx} \right)\right/ \left(2k_0 n_\| g_e \right)$. Substituting back into Eqs.~(\ref{eq:eig_ord}-\ref{eq:eig_ext}), we find that the photonic potential is proportional to $\left(\frac{d\theta}{dx}\right)^2+O(\Delta n^{-2})$; as an immediate consequence, at this order of approximation [the approximation includes to set $n_\bot\approx n_\|$ on the LHS of Eqs.~(\ref{eq:eig_ord}-\ref{eq:eig_ext})] $g_o=g_e=g$. The modulation of the beam along $z$  is now given by
\begin{equation}
 \beta_{1}^{(o)}=- \frac{ \frac{d^2 \theta}{dx^2} +2\frac{d\theta}{dx}\frac{d \ln{g}}{dx} }{ 2k_0 n_\bot } \label{eq:beta_o}
\end{equation} 
and 
\begin{equation}
\beta_{-1}^{(e)}=\frac { \frac{d^2 \theta}{dx^2} +2\frac{d\theta}{dx}\frac{d \ln{g}}{dx}} { 2k_0 n_\| },  \label{eq:beta_e}
\end{equation}
i.e., the quasi-mode undergoes periodic modulation of both amplitude and phase profiles. Since $\beta_{-1}^{(e)}\approx-\beta_{1}^{(o)}$, $\phi_o-\phi_e$ is constant in first approximation, in agreement with the initial hypothesis. According to Eq.~\eqref{eq:beta_o} it is $\beta_{-1}^{(e)}\propto \left(w^2_m k_0\right)^{-1}$, in turn providing $\frac{1}{k_0^2\left(\Delta n\right)^2} \sum_{p=1}^\infty{\frac{1}{p^2}} \frac{d \beta_p^{(e)}}{dx} \frac{d\beta_{-p}^{(e)}}{dx}\propto \left(\frac{L_B}{L_R}\right)^2$; the latter result confirms that the ratio between the beating length and the Rayleigh length is the smallness parameter in our approximated treatment.

\section{Origin of the effective photonic potential}
\label{sec:origin}

\begin{figure}
\includegraphics[width=0.5\textwidth]{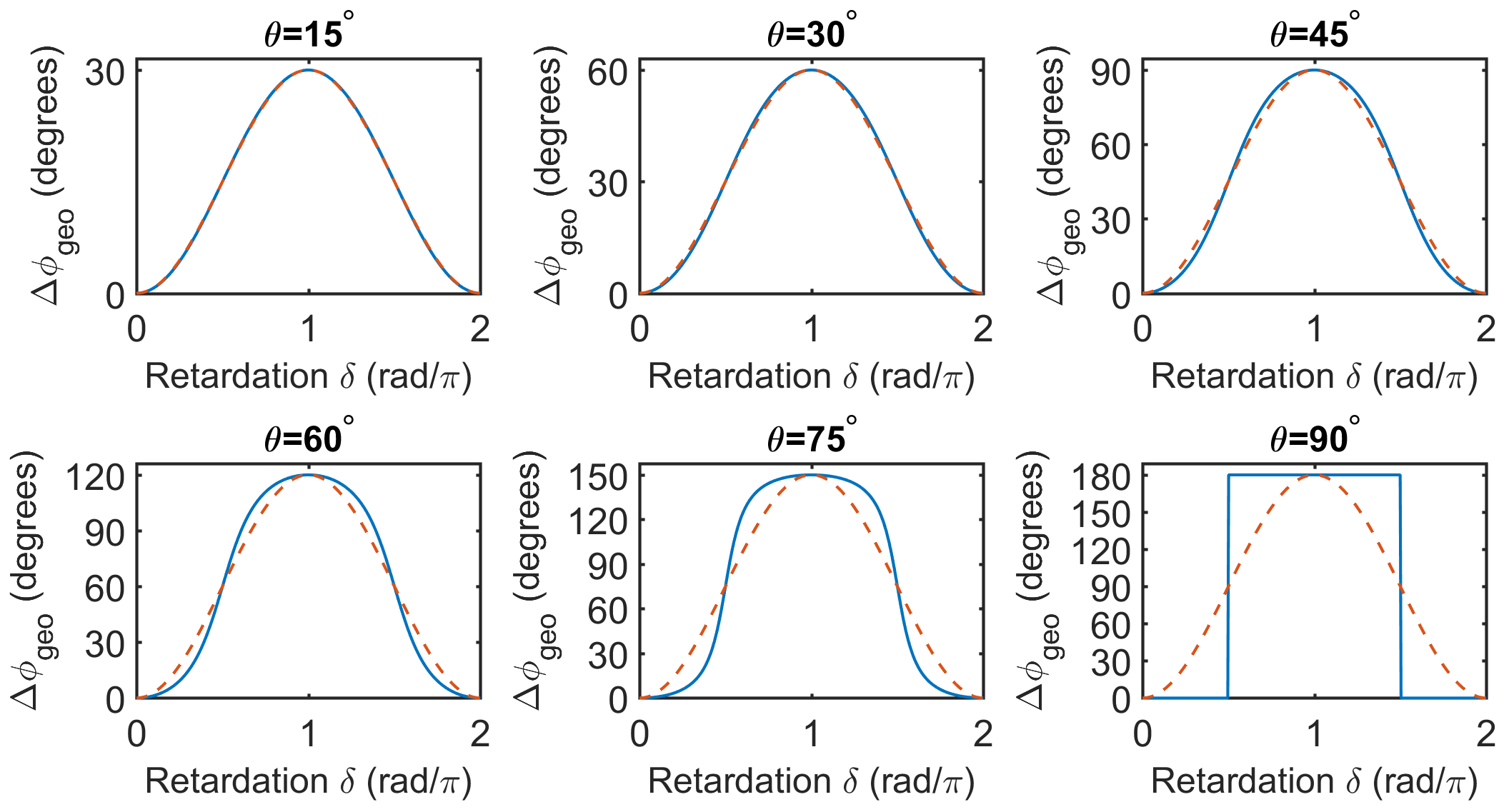}
\caption{\label{fig:geo_phase_PW} Pancharatnam-Berry phase difference $\Delta\phi_\mathrm{geo}$ versus retardation $\delta$ between two regions differing only by a relative rotation $\theta$ when spatial dispersion is neglected, i.e., in the plane wave limit. The blue solid lines are the exact results computed from Eq.~\eqref{eq:pancharatnam_phase}; the red dashed lines are a sinusoidal approximation for $\Delta\phi_\mathrm{geo}$ sharing the same peak $2\theta$ of the exact solutions.}
\end{figure}

Since we factored out the dynamic phase when introducing the fields $\psi_o$ and $\psi_e$, we can speculate that the photonic potential stems from the geometric phase. In order to prove this, in analogy with standard GRIN guides and splitting operators in numerical analysis, let us separate Eq.~\eqref{eq:maxwell_equation} in two portions, one accounting  for diffractive spreading (second derivative along $x$) and one for the inhomogeneous dielectric tensor, respectively. We guess that the transverse phase distribution from the second part is able to compensate diffraction after averaging along one beat-length $\lambda/\Delta n$. Owing to the absence of diffraction, the solutions of the second part of Eq.~\eqref{eq:maxwell_equation} correspond to Eqs.~\eqref{eq:PW}, i.e., the exact solutions in the plane wave limit. On the one hand, according to Eq.~\eqref{eq:PW}, when a purely circular polarization (either LCP or RCP) is launched, the dynamic phase accumulated in propagation is $k_0\overline{n}z$. This contribution is clearly constant in the transverse plane, thus cannot be responsible for the appearance of an $x$-dependent photonic potential. On the other hand, the terms between square brackets in Eq.~\eqref{eq:PW} provide a phase delay of geometric origin (i.e., due to variations in light polarization), which we will analyze hereafter. We adopt Pancharatnam's original approach  \cite{Pancharatnam:1956,Berry:1994} to calculate the phase difference $\Delta \phi_\mathrm{geo}$ between two different polarizations as $\Delta \phi_\mathrm{geo}= \text{arg}\left[ \bm{E}(\theta_1,z)\cdot \bm{E}^*(\theta_2,z) \right]$ (the superscript $*$ indicates the complex conjugate). For a purely RCP wave in $z=0$, Eqs.~\eqref{eq:PW} provide
\begin{equation}
  \Delta \phi_\mathrm{geo}= \text{arg}\left[ \cos^2\left(\frac{\delta}{2} \right) + \sin^2\left(\frac{\delta}{2} \right) e^{2i\left(\theta_2 -\theta_1 \right)} \right]. \label{eq:pancharatnam_phase}
\end{equation}
Fig.~\ref{fig:geo_phase_PW} graphs the geometric phase difference \eqref{eq:pancharatnam_phase} versus $\delta$ in the plane-wave limit: $\Delta \phi_\mathrm{geo}$ is periodic with  $\lambda/\Delta n$ and oscillates between the two extrema $0$ and $2\left(\theta_2-\theta_1\right)$. The oscillation shape  versus propagation $z$ depends strongly on its amplitude: it is sinusoidal for small amplitudes (the exponential can be Taylor-expanded retaining only the linear term), whereas it is flat-topped when $\theta_2-\theta_1=90^\circ$; for intermediate values of $\theta_2-\theta_1$ there is a gradual transition between these two limits.  Fig.~\ref{fig:geo_phase_PW} (red dashed lines) plots the sinusoidal approximation of Eq.~\eqref{eq:pancharatnam_phase} to illustrate the discrepancy with the exact form for various relative rotations $\theta_2-\theta_1$.  \\
Equation~\eqref{eq:pancharatnam_phase} is the key to understanding the evolution of a  finite-size beam: the beam wavefront acquires a continuous geometric phase-delay $\Delta \phi_\mathrm{geo}$, which is periodic with propagation $z$. Conversely, the amplitude of these phase-oscillations depends on the optic axis rotation between neighboring points in the transverse plane, i.e., on the $\theta$ distribution  and its derivatives across $x$. Through the Kapitza effect of light, a periodic index modulation of the form $\Delta n^2=n^2(x)-{\overline{n}}^2=f(z)W(x)$ with $f(z)=\sum_l f_l \exp{(2i\pi l z/\Lambda)}$ yields a $z$-independent effective photonic potential \cite{Alberucci:2013}
\begin{equation}
V_\mathrm{Kap}= -k_0\frac{n^2(x)-{\overline{n}}^2}{2\overline{n}} = \frac{k_0\Lambda^2}{32 \overline{n}\pi^2}\left(\sum_{l=-\infty}^\infty \frac{f_l f_{-l}}{l^2} \right) \left(\frac{d W}{dx} \right)^2. 
\end{equation} 
In essence, $V_\mathrm{Kap}$ accounts for fast scale modulations on the continuous wave (cw) component: the rapid oscillations in phase distribution entail a modulation of the transverse momentum $k_x$, the latter providing a net cumulative phase across the wavefront due to the effective kinetic energy $k_x^2/(2k_0 n_j)$ \cite{Kapitza:1951,Alberucci:2013}. \\
To further support the interpretation above, we recall that the PBP is non-transitive \cite{Berry:1987}: at any given $z$, the phase difference between two points $x_1$ and $x_2$ with rotation equal $\theta_1 =\theta (x_1)$ and $\theta_2=\theta (x_2)$, respectively, depends on the polarizations state of the beam in the whole interval $[x_1\ x_2]$ or,  otherwise stated, on its path on the Poincar\'e sphere. This is in sharp contrast with the standard dynamic phase, where the phase difference depends exclusively on initial and final state. Thus, the accumulated phase difference needs to be computed taking adjacent points in space. Fig.~\ref{fig:geo_phase_PW} shows that the profile (i.e., the Fourier coefficients $f_l$) of the periodic function $\Delta \phi_\mathrm{geo}$ along $z$ [given by Eq.~\eqref{eq:pancharatnam_phase}] markedly depends on the relative rotation $\theta=\theta_2-\theta_1$, providing distinct Kapitza potentials $V_\mathrm{Kap}$. In the limit $\left|x_2-x_1\right|\rightarrow 0$ and infinitesimally small increments of $\theta$, Eq.~\eqref{eq:pancharatnam_phase} gives $f(z)=\sin\left( \frac{2\pi \Delta n z}{\lambda}\right)$ and $W(x)=2\overline{n}\Delta n\theta(x)$. 
Such approach provides an effective $z$-invariant potential $V=\frac{1}{4 \overline{n} k_0}\left(\frac{d \theta}{d x} \right)^2$, the correct transverse profile but a factor 2 smaller than Eq.~\eqref{eq:photonic_potential}. Discrepancy arises from the fact that the Kapitza model, elaborated in Ref.~\cite{Alberucci:2013} for the scalar case, does not account for the complete spin-orbit interaction occurring in this case \cite{Bliokh:2016}. 

\section{Focusing and defocusing potentials}
\label{sec:potentials}
\begin{figure}
\includegraphics[width=0.5\textwidth]{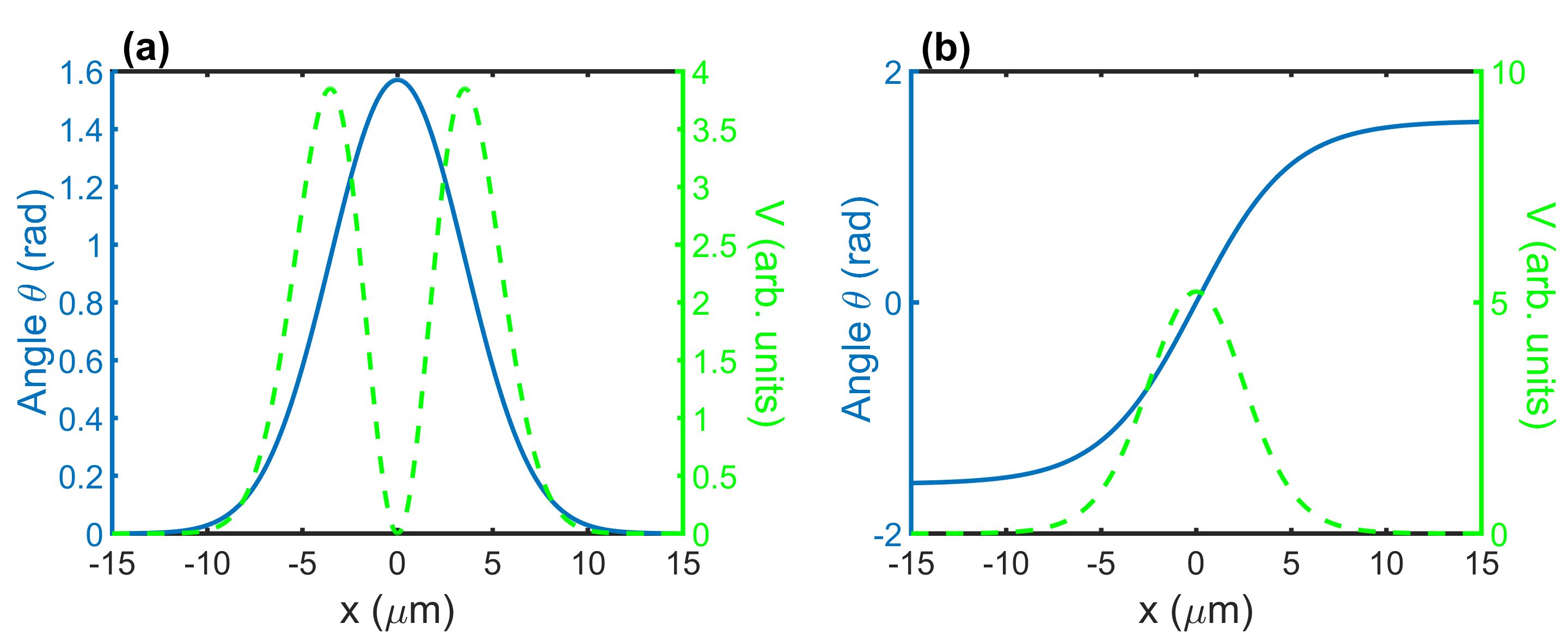}
\caption{\label{fig:mode_theory_potential} (a) Rotation angle $\theta$ (blue solid line) and corresponding photonic potential (dashed green line) versus $x$ for a Gaussian-shaped rotation of principal axes. Here $\theta_0=\pi/2$ and $w_\theta=5~\mu$m. (b) As in (a) but when $\theta$ has a hyperbolic tangent distribution. Here $L_\theta=5~\mu$m. When $\delta=\pi$, (a) and (b) behave like polarization-dependent lens and deflector, respectively.}
\end{figure}
\begin{figure}
\includegraphics[width=0.5\textwidth]{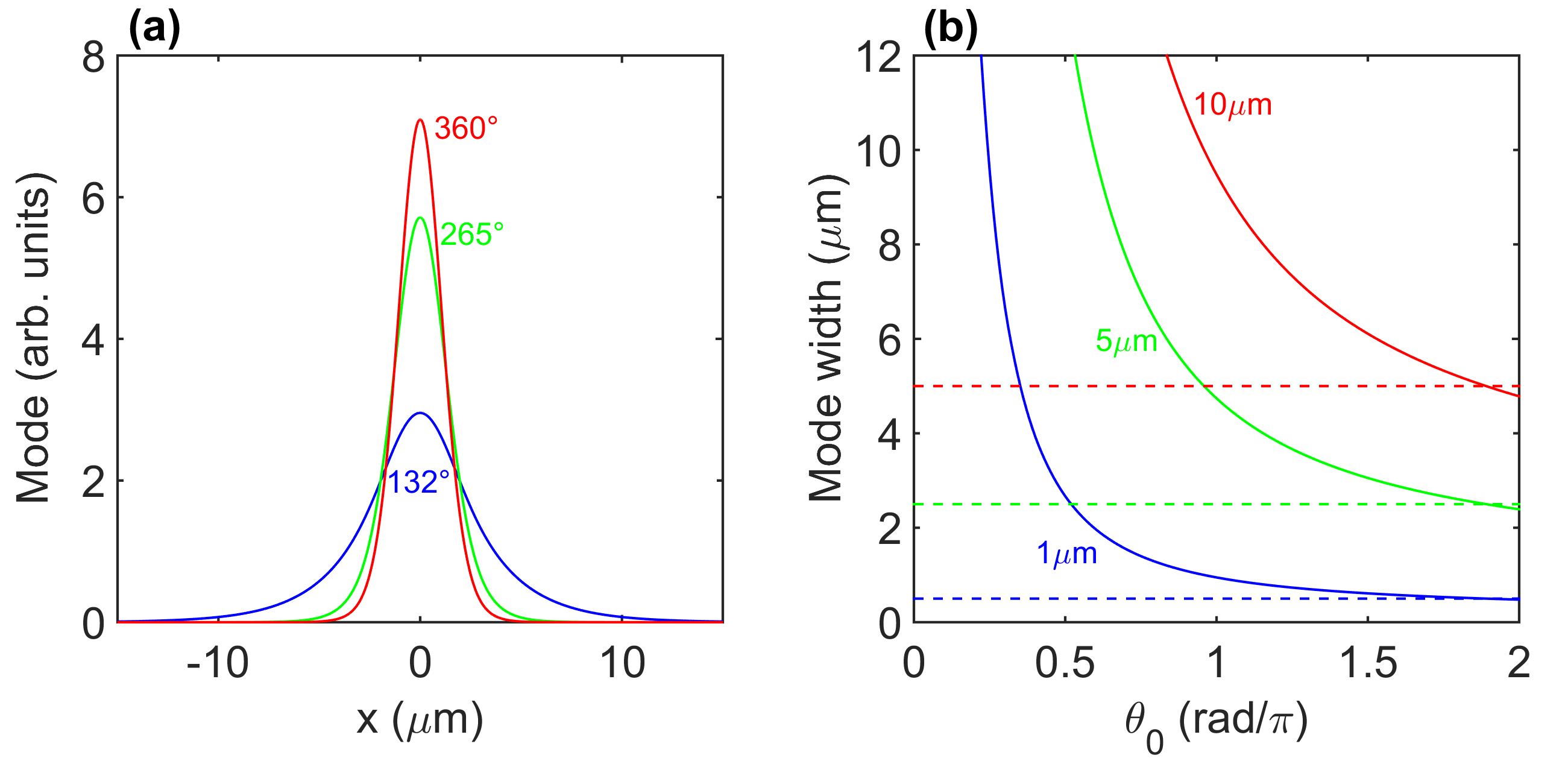}
\caption{\label{fig:mode_theory_modes} (a) Intensity profile  for a Gaussian distribution of $\theta$ and $w_\theta=5~\mu$m; each curve is labelled with the corresponding maximum rotation $\theta_0$. (b) Corresponding width of the quasi-mode versus maximum rotation angle $\theta_0$ for three different $w_\theta$, as labelled (solid lines). The dashed lines graph the half width of the angular distribution: above them the model is not accurate due to a finite overlap of the mode with the edges of the photonic potential \cite{Alberucci:2013}. Here we solved Eqs.~(\ref{eq:eig_ord}-\ref{eq:eig_ext}) considering only the term given by Eq.~\eqref{eq:photonic_potential} when computing the quasi-mode; we  assumed $n_\bot=1.5$ and $n_\|=1.7$.}
\end{figure}

According to Eq.~\eqref{eq:photonic_potential}, the shape of the effective photonic potential strongly depends on the symmetry of $\theta(x)$.
When $\theta$ is bell-shaped [solid blue line in Fig.~\ref{fig:mode_theory_potential}(a)], the potential is M-like [dashed green line in Fig.~\ref{fig:mode_theory_potential}(a)] and supports leaky modes \cite{Alberucci:2013}. Such modes can be computed with good accuracy by clipping off the  edges of the potential, i.e., approximating the M-well with a V-profile \cite{Hu:2009}. For a Gaussian distribution of $\theta$
\begin{equation}
  \theta(x)=\theta_0 e^{-\frac{x^2}{w_\theta^2}}, \label{eq:theta_gauss}
\end{equation}
Fig.~\ref{fig:mode_theory_potential}(a) plots the associated potential $V=\frac{1}{2n_j k_0}\left[ \frac{2x\theta_0}{w^2_\theta} \exp{\left(-\frac{x^2}{w_\theta^2}\right)} \right]^2$. Fig.~\ref{fig:mode_theory_modes}(a) graphs the resulting quasi-modes versus the maximum rotation $\theta_0$. The quasi-modes computed from \eqref{eq:photonic_potential} hold valid when the higher order contributions -i.e., the sum over $p$ in Eqs.~(\ref{eq:eig_ord}-\ref{eq:eig_ext})- are negligible, that is, for large enough birefringence. In Section~\ref{sec:BPM_simulations} BPM simulations will be employed to check the behavior versus the birefringence $\Delta n$. On the other hand, these results are valid as long as the mode remains confined within the central lobe of the potential [see solid and dashed lines in Fig.~\ref{fig:mode_theory_modes}(b)], regardless of the birefringence $\Delta n$ \cite{Alberucci:2013}. \\
When $\theta (x)$ is odd symmetric [solid blue line in Fig.~\ref{fig:mode_theory_potential}(b)], the photonic potential is maximum in the center (around $x=0$) and minimum at the edges: light gets repelled from the region around $x=0$ and no lateral confinement occurs. For example, if $\theta$ is given by
\begin{equation}
  \theta(x)=\theta_0 \tanh{\left(\frac{x}{L_\theta}\right)}, \label{eq:theta_tanh}
\end{equation}
the effective potential is $V=\frac{1}{2n_j k_0}\left[ \frac{\theta_0}{L_\theta} \left(1-\tanh^2{\left(\frac{x}{L_\theta}\right)} \right) \right]^2$ [see the dashed green line in Fig.~\ref{fig:mode_theory_potential}(b)].  

\section{Numerical simulations}
\label{sec:simulations}
\begin{figure}
\includegraphics[width=0.5\textwidth]{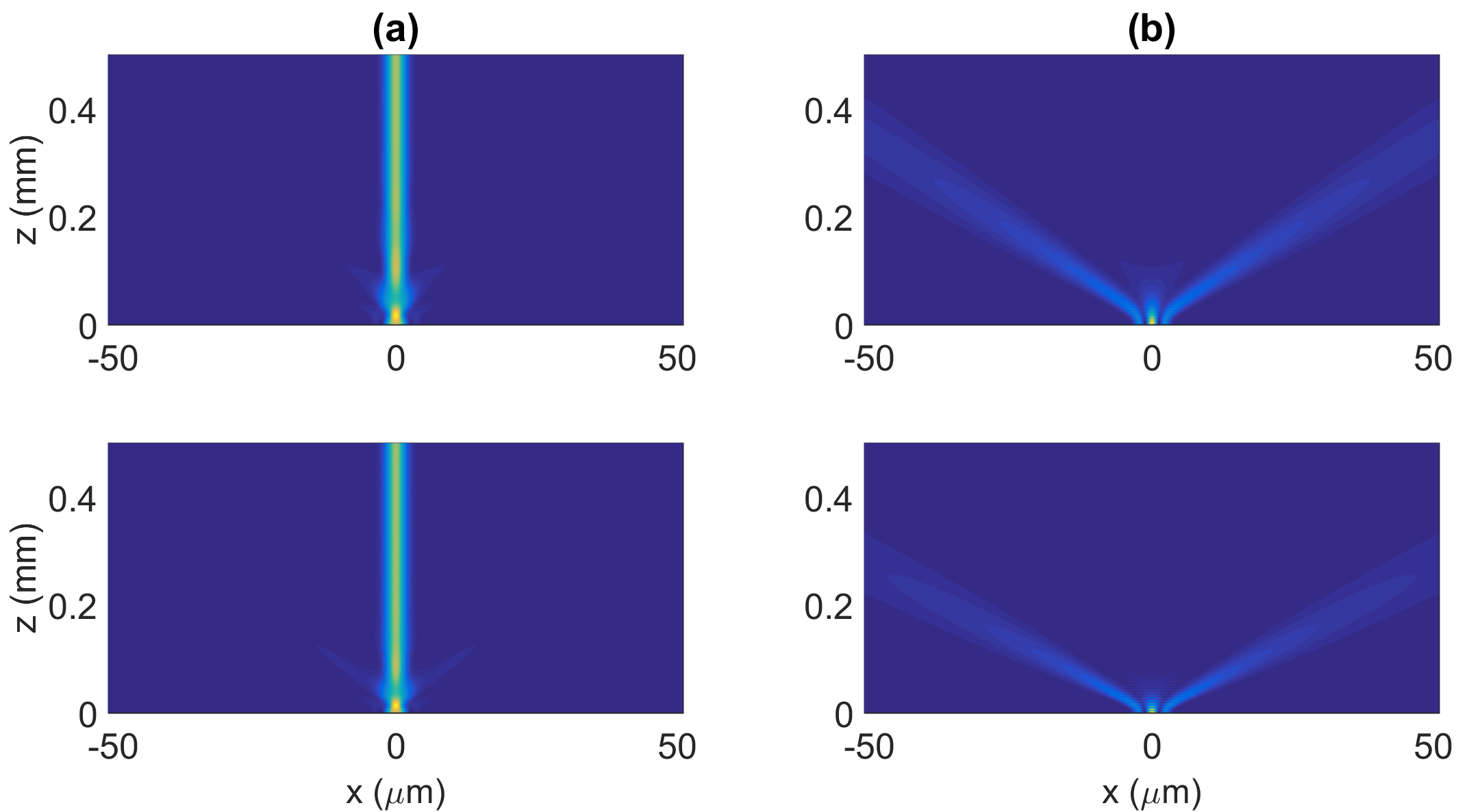}
\caption{\label{fig:BPM_focus_defocus} Light propagation in the presence of a focusing (a) or a defocusing (b) effective potential. First and second rows correspond to input wavepackets polarized along $y$ and $x$, respectively. In (a) and (b) the transverse distribution of $\theta$ is given by \eqref{eq:theta_gauss} and \eqref{eq:theta_tanh}, respectively. Here $\theta_0=360^\circ$, $w_\theta=L_\theta=5~\mu$m, $n_\bot=1.5$ and $n_\|=1.7$, the wavelength is 1064~nm. The input is a Gaussian beam of width 3~$\mu$m and flat-phasefront. }
\end{figure}
In this Section we validate the  theory developed above by means of numerical simulations. First, we use BPM simulations to verify that the photonic potential \eqref{eq:photonic_potential} describes accurately light propagation in a twisted anisotropic medium. Then, we use FDTD simulations to assess the paraxial approximation invoked in going from Eq.~\eqref{eq:maxwell_equation} to Eqs.~(\ref{eq:SVEA_ord}-\ref{eq:SVEA_ext}). With FDTD simulations we finally address the accuracy of Eq.~\eqref{eq:maxwell_equation} in lieu of the exact Maxwell's equations.
\begin{figure*}
\includegraphics[width=1\textwidth]{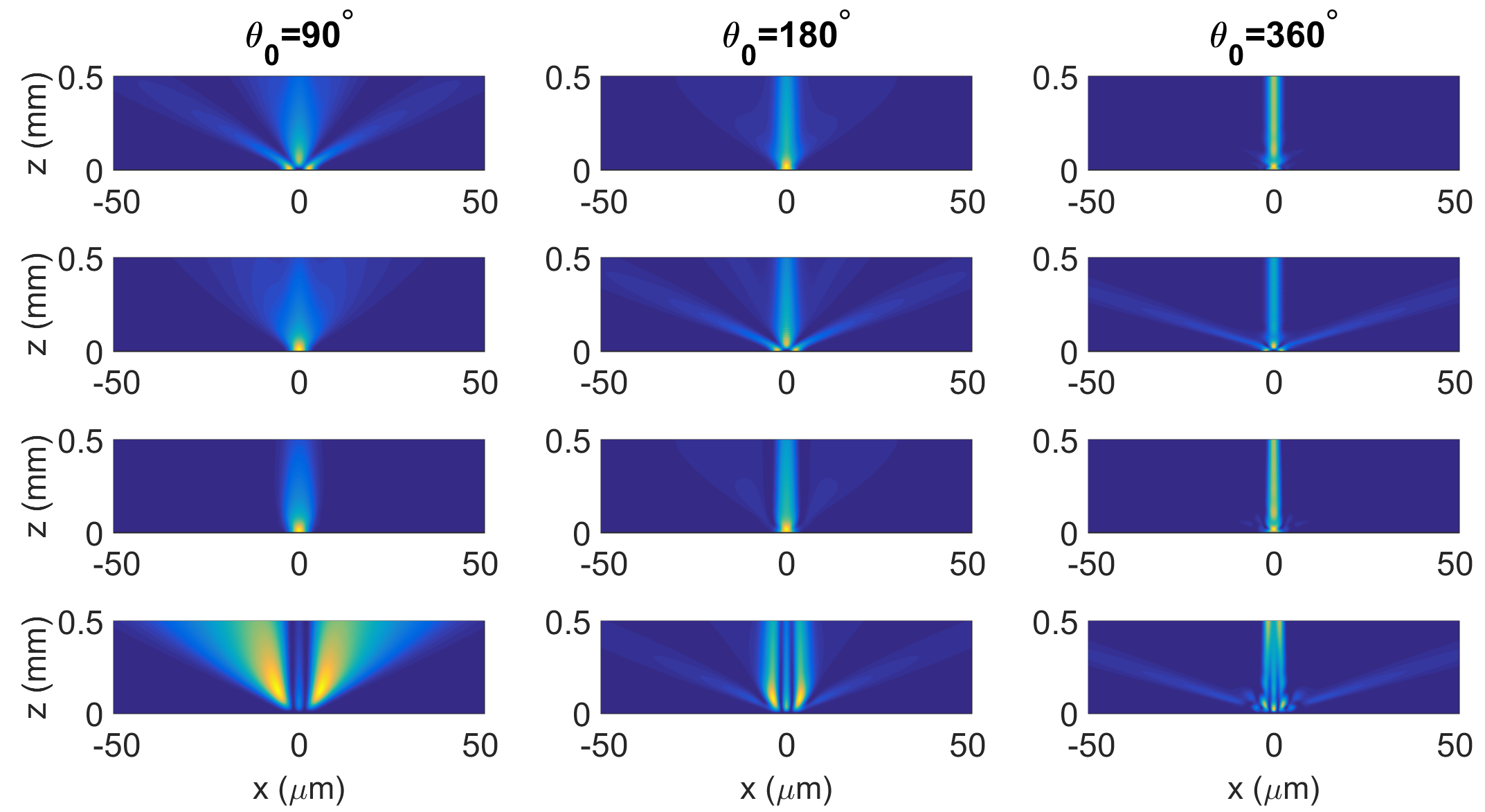}
\caption{\label{fig:BPM_vs_theta0} Intensity distribution in the plane $xz$ versus maximum rotation angle $\theta_0$, computed with BPM simulations when $\theta$ is Gaussian with $w_\theta=5~\mu$m, see Eq.~\eqref{eq:theta_gauss}; here the refractive indices are $n_\bot=1.5$ and $n_\|=1.7$ and a $y$-polarized Gaussian beam of waist $3~\mu$m is launched at the input. In the first and second rows $\left|\psi_e\right|^2$ and $\left|\psi_o\right|^2$ are plotted, respectively. In the third and fourth rows the corresponding fields  $\left|E_y\right|^2$ and $\left|E_x\right|^2$ are graphed in the laboratory coordinate system, respectively.}
\end{figure*}

\subsection{BPM simulations}
\label{sec:BPM_simulations}
\begin{figure*}
\includegraphics[width=1\textwidth]{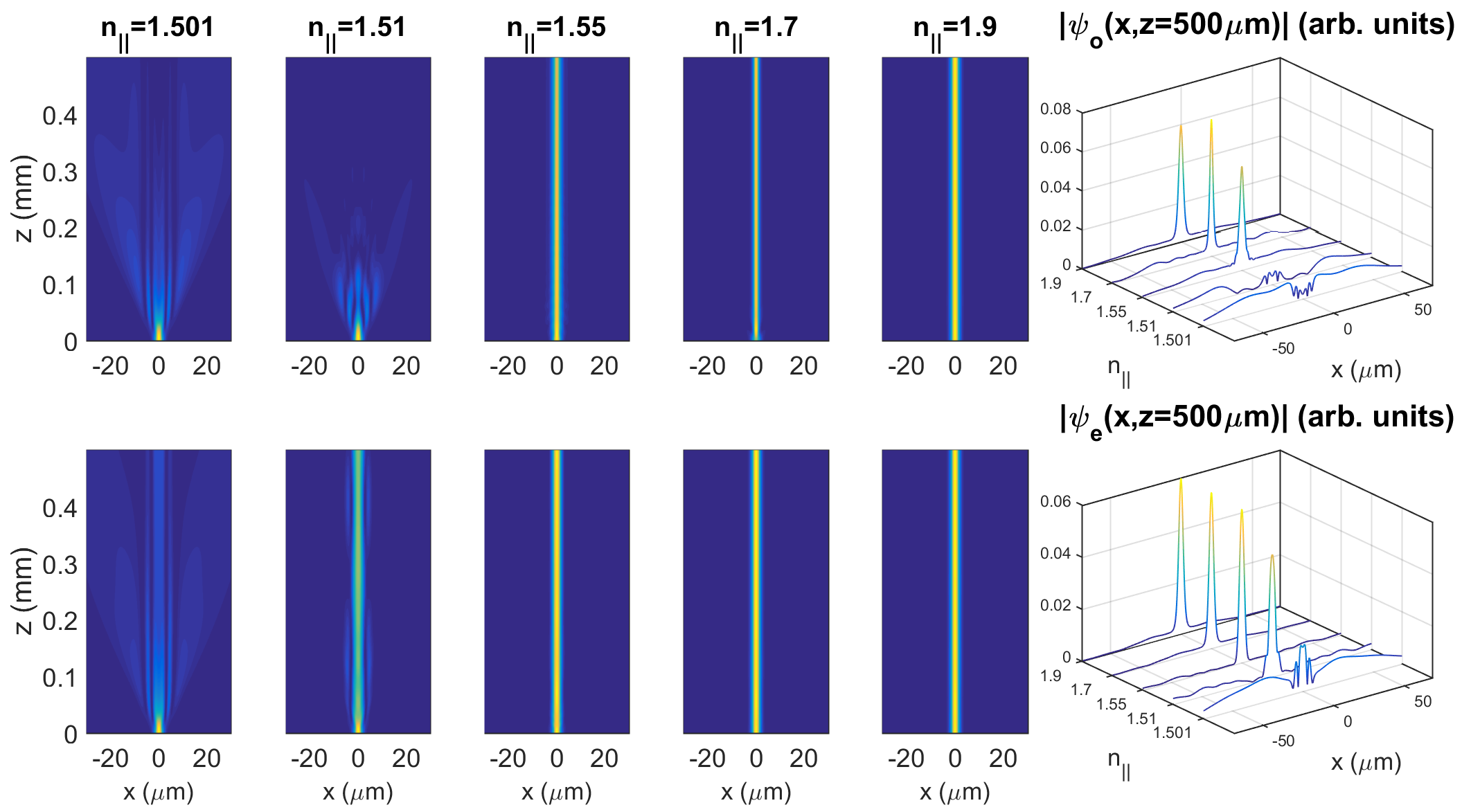}
\caption{\label{fig:BPM_modes} Columns 1 to 5: Intensity evolution in the plane $xz$ when the mode plotted in Fig.~\ref{fig:mode_theory_modes} is launched at the input $z=0$, for different values of $n_\|$ as labelled (the corresponding birefringence $\Delta n$ is 0.001, 0.01, 0.05, 0.2 and 0.4, from left to right respectively). Top and bottom rows correspond to local ordinary and extraordinary polarizations, respectively. Column 6: modulus of the field $z=500~\mu$m versus $x$ and the refractive index $n_\|$. Here $\theta$ is Gaussian with $\theta_0=360^\circ$ and $w_\theta=5~\mu$m, $n_\bot=1.5$.}
\end{figure*}

The presence of the effective potential given by Eq.~\eqref{eq:photonic_potential} can be tested by simulating Eqs.~(\ref{eq:SVEA_ord}-\ref{eq:SVEA_ext}), valid in the paraxial limit. To simulate light evolution in the rotated reference system, we use a BPM code based on operator splitting and the Crank-Nicolson algorithm for  diffraction. \\
First, we check that twisted anisotropic materials exhibit a quasi-isotropic response for propagation lengths much longer than the Rayleigh distance and large enough birefringence. The results are shown in Fig.~\ref{fig:BPM_focus_defocus} for even [Eq.~\eqref{eq:theta_gauss} and Fig.~\ref{fig:mode_theory_potential}(a)] and odd symmetry potentials [Eq.~\eqref{eq:theta_tanh} and Fig.~\ref{fig:mode_theory_potential}(b)], with $n_\bot=1.5$ and $\Delta n=0.2$, a birefringence large enough to ensure that Eq.~\eqref{eq:photonic_potential} is a good approximation to the exact photonic potential. For $\delta=\pi$, the even and odd distributions correspond to a polarization-dependent lens \cite{Roux:2006} and a polarization-dependent deflector based on the spin-Hall effect \cite{Li:2013}, respectively. As predicted by our theory, in both cases the electromagnetic propagation is nearly independent of the input polarization. If the $\theta$ distribution is bell-shaped, diffractive spreading is counteracted and light is laterally trapped, i.e., waveguiding takes place. If $\theta$ is odd, light is repelled from the symmetry axis $x=0$. Summarizing, Fig.~\ref{fig:BPM_focus_defocus} confirms qualitatively that light is subject to an effective potential given by \eqref{eq:photonic_potential}.  \\
Figure~\ref{fig:BPM_vs_theta0} displays light propagation for a Gaussian distribution of $\theta$ [see Eq.~\eqref{eq:theta_gauss} and Fig.~\ref{fig:mode_theory_potential}(a)] versus the maximum rotation $\theta_0$, with $n_\bot=1.5$ and $\Delta n=0.2$. In agreement with Eq.~\eqref{eq:photonic_potential}, as $\theta_0$ increases the beams undergo a stronger transverse confinement due to the M-shaped potential. At the same time, the figure shows how a purely $y$-polarized input beam couples a fraction of its power into the $x$-polarization owing to the pointwise rotation of the optic axis. Regardless of $\theta_0$, strong coupling to radiation is observed due to the small spatial overlap between a  linearly polarized Gaussian beam and the leaky modes. \\
Next, we check that our analytical model is in quantitative agreement with the actual solutions, as well. It is also important to study how electromagnetic propagation depends on the birefringence $\Delta n$. To that extent, at the input we injected the quasi-modes computed from the eigenvalue problem, Eqs.~(\ref{eq:ORD}-\ref{eq:EXT}), in the limit of a large birefringence $\Delta n$, i.e., when only the action of the photonic potential provided by Eq.~\eqref{eq:photonic_potential} is considered, see Fig.~\ref{fig:mode_theory_modes}. By means of Eqs.~(\ref{eq:beta_o}-\ref{eq:beta_e}), our theory predicts that the second term between square brackets in Eqs.~(\ref{eq:ORD}-\ref{eq:EXT}) goes as $(\Delta n)^{-2}$, and thus can be neglected for large enough birefringence. Figure~\ref{fig:BPM_modes} illustrates the quasi-mode evolution in the medium for $n_\bot=1.5$ and varying the birefringence. For $\Delta n>0.1$, guiding  is clearly observed for both $\psi_o$ and $\psi_e$: this implies that, for $\Delta n> 0.1$, higher order terms in Eqs.~(\ref{eq:ORD}-\ref{eq:EXT}) can be neglected. For larger birefringence the generation of lateral wings is observed. In this limit the difference between $n_\bot$ and $n_\|$ is large enough to be appreciable even on the LHS of Eqs.~(\ref{eq:ORD}-\ref{eq:EXT}), yielding a slightly anisotropic response on long scales (with respect to the beat length $\lambda/\Delta n$). For birefringence $<0.1$, higher order terms in Eqs.~(\ref{eq:ORD}-\ref{eq:EXT}) are relevant and the differences between the exact photonic potential and its approximation given by Eq.~\eqref{eq:photonic_potential} cannot be neglected anymore: as a net result, guidance is lost and a polarization-dependent behavior arises. In particular, the beam remains partially trapped for birefringence down to $\Delta n\approx 0.03$ and $\Delta n\approx 0.01$ for ordinary and extraordinary inputs, respectively. Finally, the power coupled to guided modes strongly depends on the overlap between the input and the quasi-mode, as shown by a direct comparison between Figs.~\ref{fig:BPM_modes} and \ref{fig:BPM_vs_theta0}. We note that the overlap does not depend solely on the spatial distribution of the optical wavepacket, but also on its local polarization: in other words, the quasi-modes are structured light.

\subsection{FDTD simulations}
\begin{figure*}
\includegraphics[width=1\textwidth]{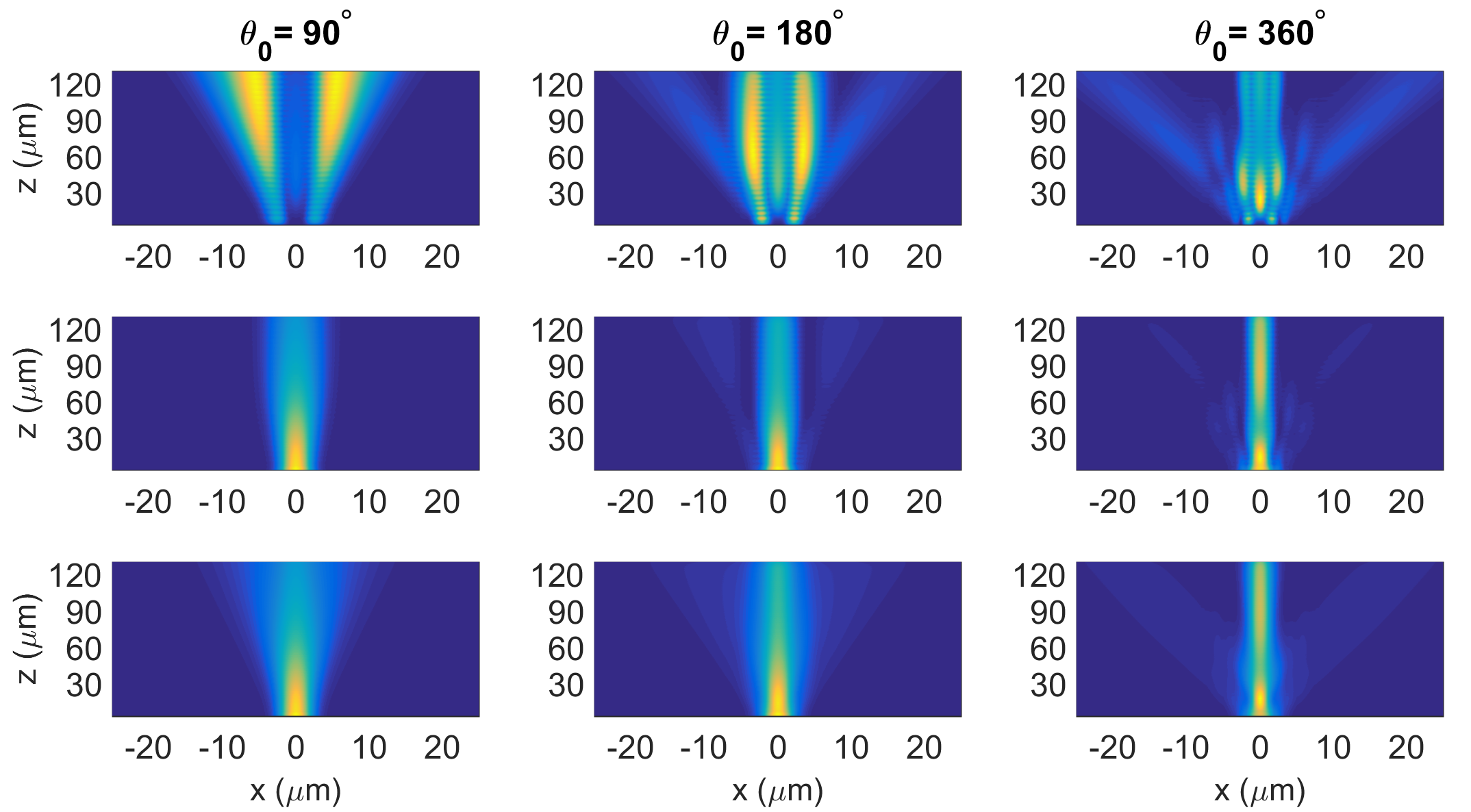}
\caption{\label{fig:FDTD_theta0} FDTD simulations versus $\theta_0$. Evolution of $\left|E_x\right|^2$ (first row), $\left|E_y\right|^2$ (second row) and the overall intensity $\left|E_x\right|^2+\left|E_y\right|^2$ (third row) in the plane $xz$. The input beam is linearly polarized along $y$. } 
\end{figure*}

We now check the previous results by FDTD simulations, solving directly the Maxwell's equations in the time domain and without the paraxial approximation. We used the open-source code MEEP \cite{Oskooi:2010}, taking a continuous-wave excitation at the vacuum wavelength $\lambda=1~\mu $m, although our findings hold valid regardless of the frequency. The source was a Gaussian-shaped collection of dipoles, of width $3~\mu $m across $x$, infinitesimally narrow along $z$ and centered in $x=z=0$. The emitted radiation from the source was uniformly linearly polarized. The inhomogeneous uniaxial medium started in $z=2~\mu $m. \\ 
\begin{figure}
\includegraphics[width=0.5\textwidth]{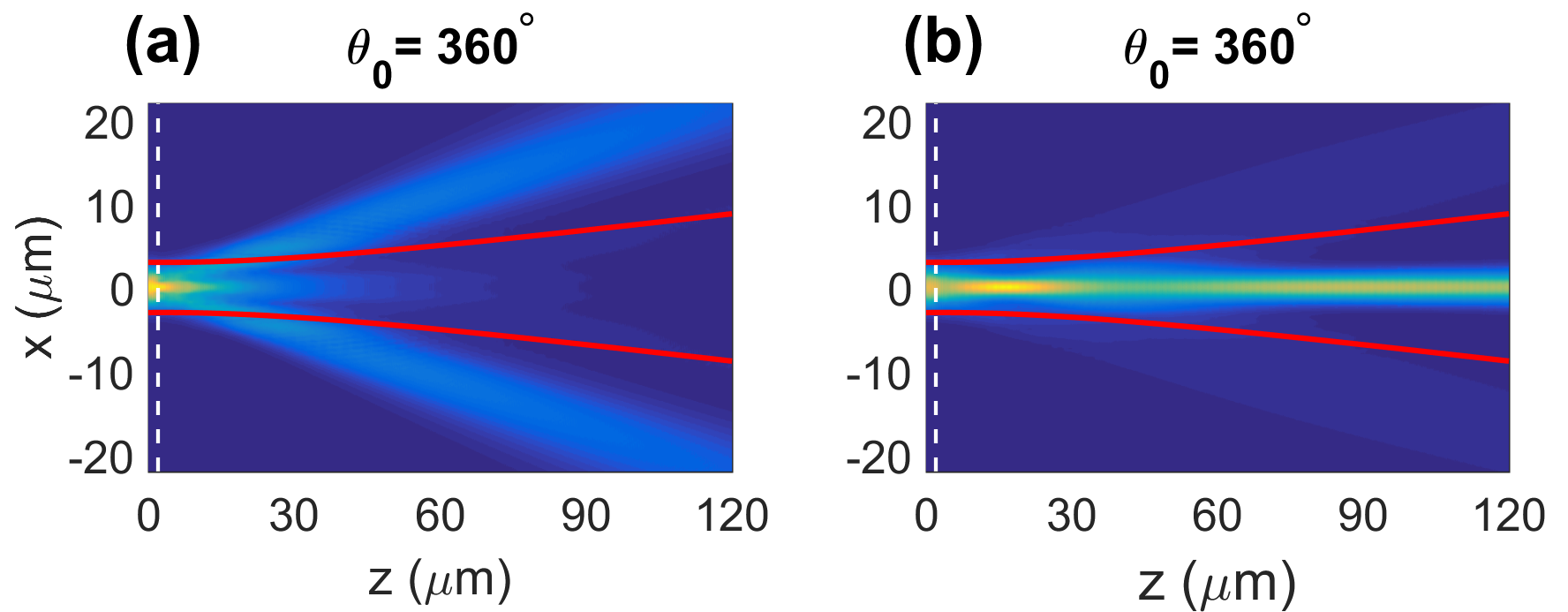}
\caption{\label{fig:FDTD_basic} FDTD simulations of the intensity distribution for defocusing [panel a, $\theta$ obeying Eq.~\eqref{eq:theta_gauss}] and focusing [panel b, $\theta$ distributions as provided by Eq.~\eqref{eq:theta_tanh}] effective potential. The input beam is linearly polarized along $y$, $n_\|=1.7$, $w_\theta=L_\theta=5~\mu$m and $\theta_0=360^\circ$. The white dashed line indicates the input interface, the red solid lines show beam diffraction for $\theta_0=0^\circ$. } 
\end{figure}
We first verify that the bell-shaped and odd $\theta$ distributions correspond to a trapping and a repulsive dynamics, respectively. We will use $n_\bot=1.5$ and $n_\|=1.7$ (thus $\Delta n=0.2$) until otherwise specified. Fig.~\ref{fig:FDTD_basic}(a) illustrates the wavepacket evolution when the rotation angle is given by \eqref{eq:theta_tanh} [see Fig.~\ref{fig:mode_theory_potential}(b)]: light is expelled from the central region. Fig.~\ref{fig:FDTD_basic}(b) graphs the confining case corresponding to \eqref{eq:theta_gauss} [see Fig.~\ref{fig:mode_theory_potential}(a)]. In both cases, the FDTD results are in excellent agreement with the BPM simulations in Fig.~\ref{fig:BPM_focus_defocus}.  \\
We then concentrate on the trend of the guiding effect [$\theta$ satisfying Eq.~\eqref{eq:theta_gauss}] versus the maximum rotation angle $\theta_0$, see Fig.~\ref{fig:FDTD_theta0}. The confinement is strongly enhanced as the maximum rotation increases, reaching a good degree of confinement for $\theta_0=360^\circ$. A comparison with Fig.~\ref{fig:BPM_vs_theta0} is in excellent agreement with the BPM results.\\
\begin{figure}
\includegraphics[width=0.5\textwidth]{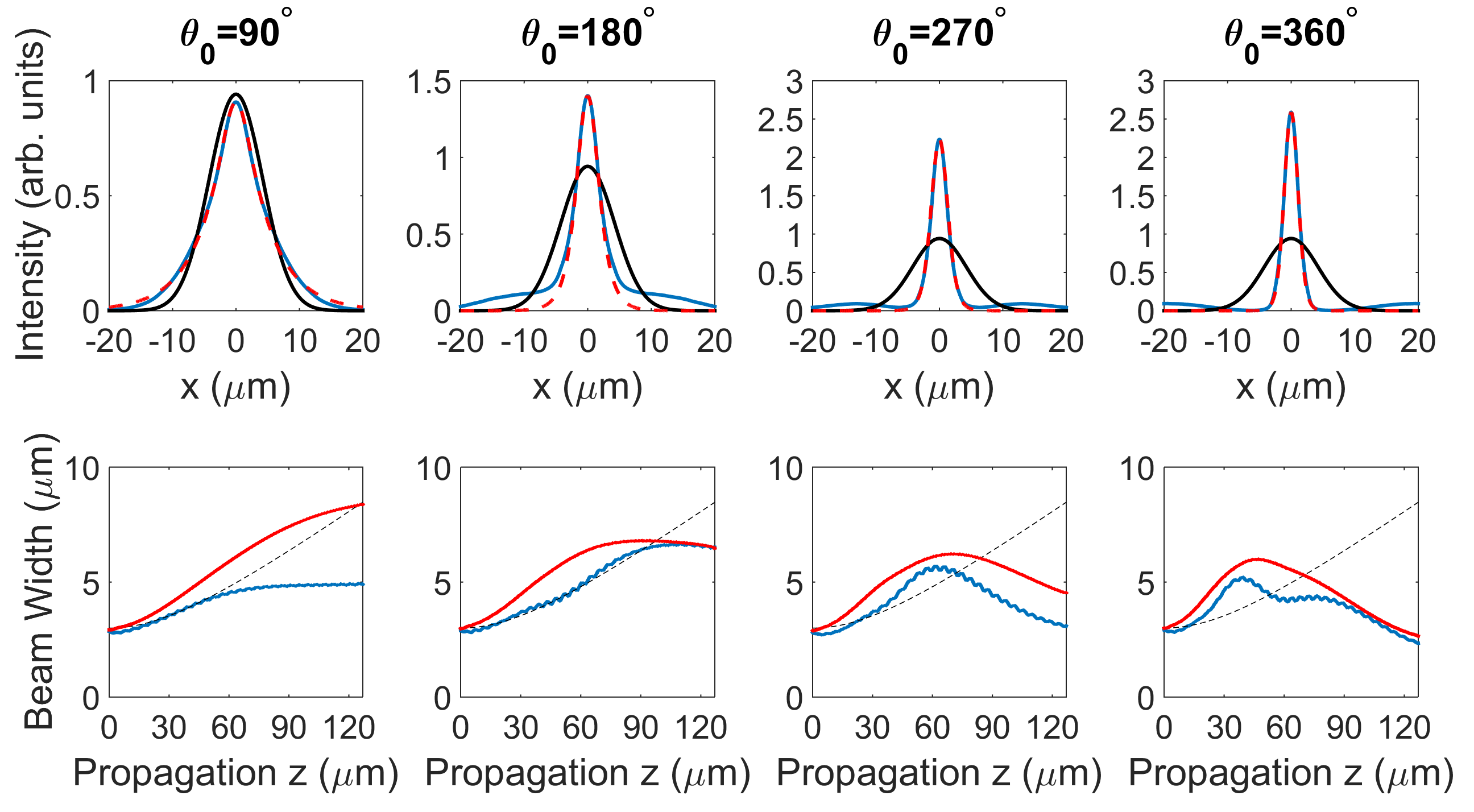}
\caption{\label{fig:FDTD_theory_comparison} Direct comparison between analytic theory and FDTD simulations when the input is linearly polarized along $y$, $\theta$ has is Gaussian  and $w_\theta=5~\mu $m. Upper panels: overall intensity cross-sections calculated at $z=120~\mu $m for a Gaussian wavepacket undergoing diffraction when $\theta_0=0^\circ$ (solid black lines) and the dielectric tensor is transversely rotated (solid blue lines). The red dashed  lines are the corresponding quasi-modes from Eqs.~(\ref{eq:eig_ord}-\ref{eq:eig_ext}). Lower panels: wavepacket width versus propagation distance $z$ using $\Pi_{E_y}$ (blue solid line) and $\Pi_I$ (red solid lines) for the probability densities. The black dashed line corresponds to a diffracting Gaussian beam in the limit $\theta_0=0^\circ$. Here $L=10~\mu$m.}
\end{figure}
Quantitative details on the evolution of the input wavepacket are gathered by computing its width $w_\eta =2\sqrt{\int_{-L}^{L}{x^2 \Pi_\eta (x) dx}}\ (\eta=I,E_j)$ versus propagation $z$, with $\Pi_I=I/\int{Idx}$ and $\Pi_{E_j}=\left|E_j\right|^2/\int{\left|E_j\right|^2 dx}$ ($j=o,e$) the probability densities for intensity and electric fields, respectively. The integrals are limited to the domain $[-L\ L]$ in order to get rid of the diffracting components generated at the input interface and then radiated out. Such densities allow evaluating how power shares between the two orthogonal polarizations. The second row in Fig.~\ref{fig:FDTD_theory_comparison} shows beam size versus $z$ and versus the maximum rotation $\theta_0$, respectively. The beam width $w_I$  initially broadens due to the strong radiation emitted at the input interface and the coupling into the guide. Over longer propagation distances the wavepacket narrows down and stabilizes to a width corresponding to the leaky mode provided by Eq.~\eqref{eq:photonic_potential}. We stress that, after the radiation excited at the input interface fades out, another kind of coupling to radiation occurs, inherent to leaky modes in M-shaped waveguides \cite{Hu:2009}. The corresponding losses strongly depend on $\theta_0$, becoming negligible after several Rayleigh distances for modes narrower than the guiding core of the M-guide. The two different contributions to radiation can be discriminated more easily by BPM simulations, as in that case the input can be tailored to match the effective mode. The role of the overlap in $z=0$ can be evaluated by comparing Fig.~\ref{fig:BPM_vs_theta0} and \ref{fig:BPM_modes} for $n_\|=1.7$ (corresponding to $\Delta n=0.2$) and $\theta_0=360^\circ$: when a Gaussian beam is launched, at the input interface two additional beams are emitted, tilted to left and right of the impinging wave vector by the same angle. These side lobes are strongly dampened when the quasi-mode is launched. The radiation loss in the bulk, related with the leaky nature of the quasi-mode, can be better appreciated for $\theta_0=180^\circ$ and $n_\|=1.7$. In fact, once radiation from the interface has moved away from the central region, the amplitude of the central lobe slightly decreases with $z$, both in BPM (Fig.~\ref{fig:BPM_vs_theta0}) and in FDTD simulations (Fig.~\ref{fig:FDTD_theta0}). \\
\begin{figure}
\includegraphics[width=0.5\textwidth]{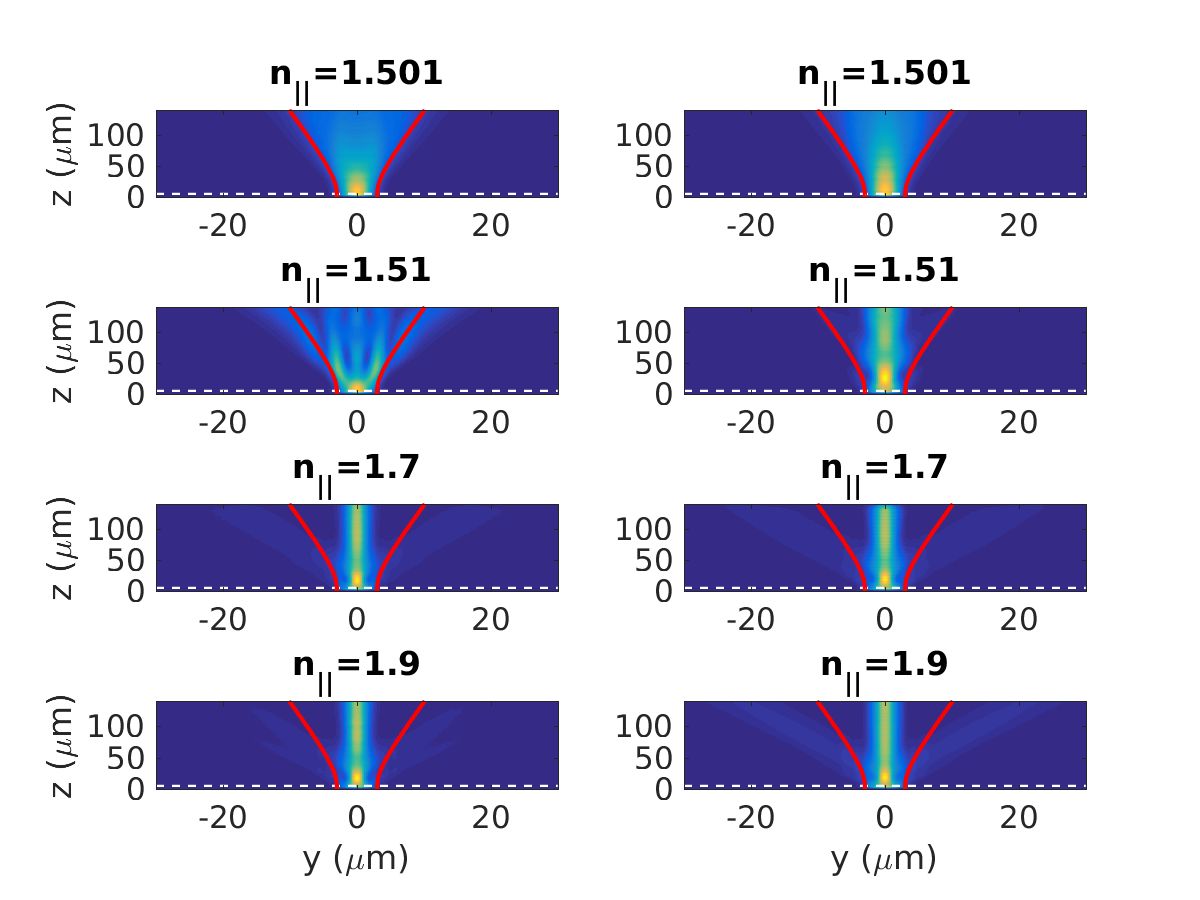}
\caption{\label{fig:FDTD_birefringence} FDTD simulations versus birefringence for a fixed  $\theta$ profile. The beam is linearly polarized along $x$ (left column) and $y$ (right column). Birefringence $\Delta n$ is 0.001, 0.01, 0.2 and 0.4, from top to bottom row respectively. The red solid lines show the diffracting case, the dashed white lines correspond to the initial section of the anisotropic medium. Here $n_\bot=1.5$, $\theta_0=360^\circ$ and $w_\theta=5~\mu$m.} 
\end{figure}
We finally investigate light propagation versus input polarization and birefringence, the results shown in Fig.~\ref{fig:FDTD_birefringence}. For $\Delta n>0.2$, the  propagation is almost independent from the input polarization; at lower birefringence, the propagation is polarization-dependent. These findings match the BPM results, see e.g., Fig.~\ref{fig:BPM_modes}. When $n_\|=1.501$ and $n_\|=1.51$ (corresponding to $\Delta n=0.001$ and $\Delta n=0.01$, respectively), light propagation is polarization-dependent both in BPM and FDTD simulations, owing to the action of the higher order contributions, the latter modifying the shape of
the photonic potential with respect to the approximated formula Eq.~\eqref{eq:photonic_potential}. In both cases, light trapping is stronger for input beams polarized along $y$ than along $x$. Specifically, for $n_\|=1.51$ input wavepackets polarized along $y$ are confined while the orthogonal polarization spreads out, whereas for $n_\|=1.501$ wavepacket widens irrespectively of the input polarization. Noteworthy, when BPM (Fig.~\ref{fig:BPM_modes}) and FDTD (Fig.~\ref{fig:FDTD_birefringence}) are compared, it is important to recall that propagation lengths in the former are much larger than in the latter.

\section{Non-leaky waveguides}
\label{sec:nonleaky}

\begin{figure}
\includegraphics[width=0.5\textwidth]{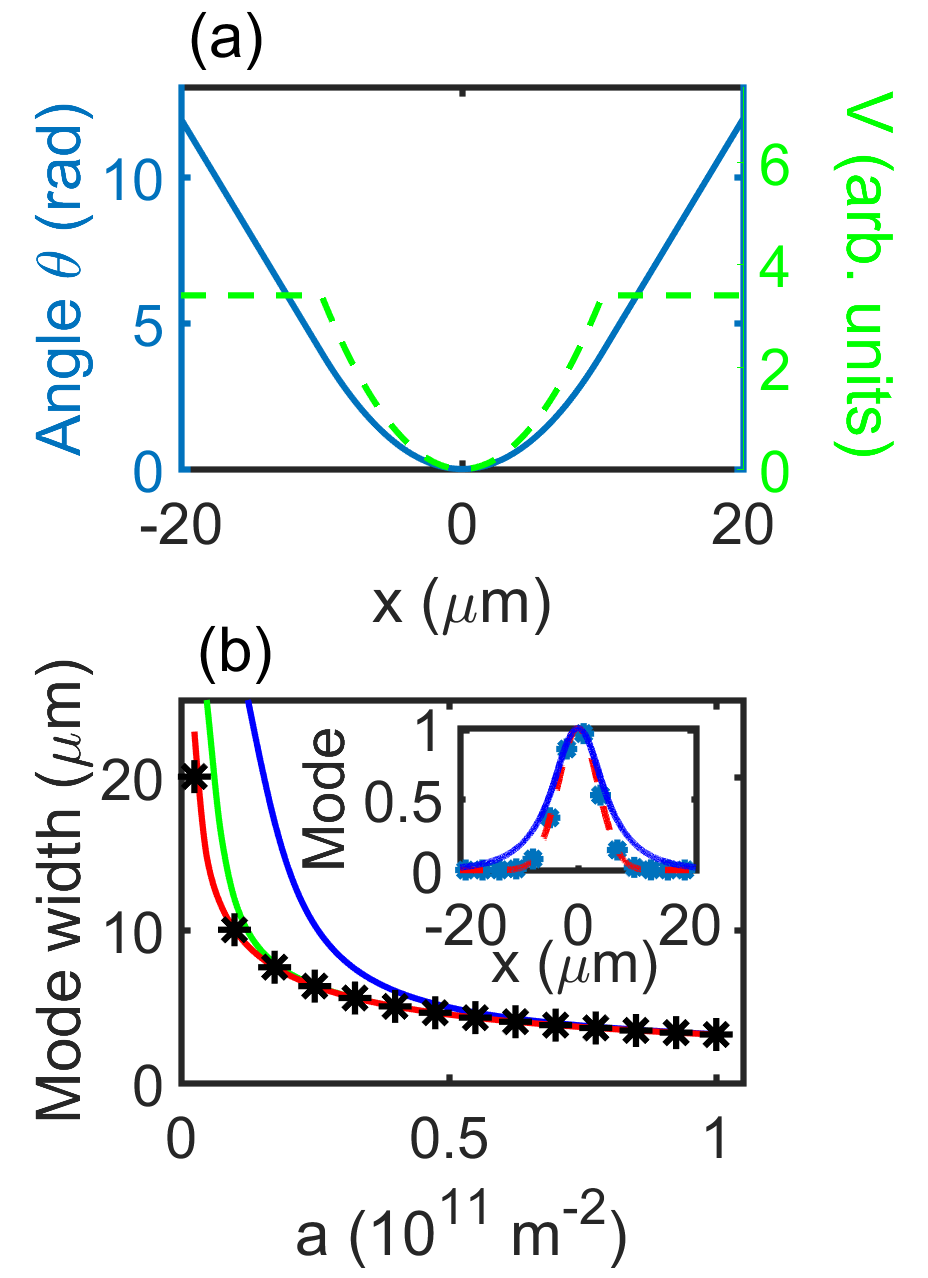}
\caption{\label{fig:guide_noleaky} V-shaped PBP waveguide. (a) Rotation angle $\theta$ (blue solid line) and corresponding potential $V$ from Eq.~\eqref{eq:potential_Vshaped} (green dashed line) for $x_0=10~\mu $m and $a=4\times 10^{10}~$m$^{-2}$. (b) Mode width versus $a$ for (top to bottom curves) $x_0=5~\mu $m (blue line), $10~\mu $m (green) and $20~\mu $m (red). The stars graph the width $w_G$ of the Gaussian fundamental mode of the parabolic potential in the limit $x_0\rightarrow\infty$. Inset:  mode intensity profile when $a=2\times 10^{10}~$m$^{-2}$ for $x_0=5~\mu $m (blue solid line) and  $x_0=20~\mu $m (red solid line). Blue stars represent the Gaussian mode  when $x_0\rightarrow\infty$. } 
\end{figure}

\begin{figure}
\includegraphics[width=0.5\textwidth]{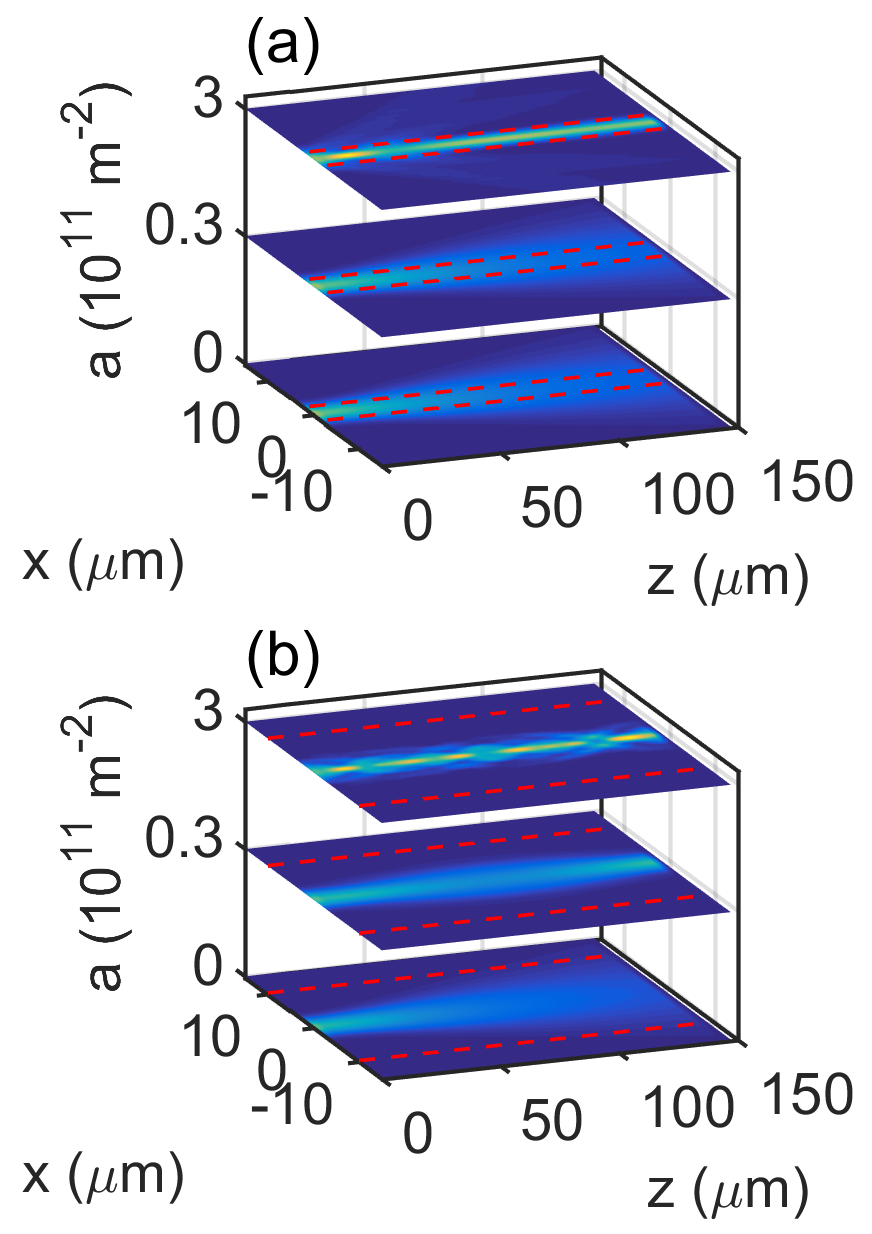}
\caption{\label{fig:guide_noleaky_FDTD} V-shaped PBP waveguide. (a-b) FDTD simulations in the plane $xz$ for $a=0,\ 3\times 10^{10}$ and $3\times 10^{11}~$m$^{-2}$, respectively (slices from bottom to top) when (a) $x_0=2~\mu $m  and (b) $x_0=10~\mu $m. The red dashed lines correspond to the straight lines $x=\pm \left|x_0\right|$, i.e., the waveguide edges. } 
\end{figure}

Equation \eqref{eq:photonic_potential} allows tailoring the profile $\theta(x)$ of the continuous rotation to yield a bell-shaped effective well able to support non-leaky guided modes. Since the lowest value of the potential is zero, for a non-leaky electromagnetic waveguide the derivative $d\theta/dx$  must vanish in $x=0$, corresponding to a local extremum of $\theta$. Therefore, $\theta(x)$ needs to be an even function. The condition $V(x)>0$ for $\left|x\right|\rightarrow \infty$ must be satisfied to get evanescent waves beyond the waveguide edges. Moreover, a realistic waveguide must be (transversely) finite; hence, we have to impose a linear trend $\theta(x)=\alpha \left( \left| x \right|- x^\diamondsuit \right) $) for $\left|x\right|>x_0$, with $x_0$ linked to the width of the guide. Finally, for $\left|x\right|<x_0$ we need to select a convex even function $\theta$ so that the potential \eqref{eq:photonic_potential} has a relative minimum in $x=0$.  The simplest choice is a parabolic profile of the form $ax^2$, such that the resulting photonic potential $V$ is
\begin{align}  \label{eq:potential_Vshaped}
   V(x) =
  \begin{cases}
    \frac{2a^2x^2}{n_j k_0}      & \quad \mathrm{for } \left|x\right|\leq x_0, \\
    \frac{2a^2x_0^2}{n_j k_0}    & \quad \mathrm{for } \left|x\right|>x_0,     \\
  \end{cases}
\end{align}
where we satisfied the continuity conditions on $\theta(x)$ and $d\theta/dx$ by setting $x^\diamondsuit=x_0/2$ and $\alpha=2ax_0$, respectively. An example is graphed in Fig.~\ref{fig:guide_noleaky}(a).\\
In the limit $x_0\rightarrow\infty$ the photonic potential is purely parabolic, thus supporting an infinite number of guided modes, the fundamental one being Gaussian with field proportional to $e^{-x^2/w_G^2}$ and width $w_G=1/\sqrt{a}$. The fundamental mode of the PBP waveguide defined by Eq.~\eqref{eq:potential_Vshaped} corresponds to such solution for widths $w_G$ much smaller than $x_0$, as confirmed in Fig.~\ref{fig:guide_noleaky}(b) graphing the exact modes of the potential Eq.~\eqref{eq:potential_Vshaped}. 
We analyze the beam propagation in such PBP waveguide by means of FDTD simulations: a synopsis of the results is available in Fig.~\ref{fig:guide_noleaky_FDTD}(a-b). As the coefficient $a$ increases starting from zero, the wavepacket undergoes transverse confinement. In agreement with Fig.~\ref{fig:guide_noleaky}(b), narrower waveguides (i.e., smaller $x_0$) yield lower field confinement.      

\section{Conclusions and Perspectives}
\label{sec:conclusions}
In this Paper we examined the propagation of electromagnetic waves in inhomogeneously anisotropic media, locally twisted in the transverse plane but invariant in propagation. First, we expanded the results reported in Ref.~\cite{Alberucci:2016}, providing a deeper insight on the physical principles and extending the theoretical computations. As a matter of fact, we showed that light propagates under the influence of an effective photonic potential due to the spin-orbit interaction, which originates from the periodic modulation of the Pancharatnam-Berry phase versus propagation via the Kapitza effect. Noteworthy, this photonic potential features a quasi-isotropic behavior, despite the locally anisotropic response of the material. The attracting or repelling character of the potential  strictly depends on the symmetry of the optic axis rotation: in particular, guiding structures correspond to bell-shaped distributions. Due to the inherent power exchange between the ordinary and the extraordinary components, the system does not support $z$-invariant modes \cite{Bliokh:2007} but  quasi-periodic modes, in agreement with the Floquet theorem. While in general the guided modes are leaky owing to an M-shaped potential, we designed a distribution yielding a V-shaped potential and truly guided modes. \\
With respect to Ref.~\cite{Alberucci:2016}, all of these results have been (qualitatively and quantitatively) validated by means of BPM and FDTD numerical simulations. In particular, here we used for the first time a BPM written in the rotated framework, specifically developed for this problem. We also demonstrated a very good agreement between BPM and FDTD: this is a very important result given that BPM codes are much less demanding than FDTD from the computational point of view. Our new BPM code will be thus relevant for the numerical analysis of q-plates and similar devices, nowadays mainly described by means of simple plane-wave models. Both the numerical techniques confirmed that our theory provides a very good approximation of the confined mode for a vast range of parameters. With respect to the latter statement, we also explored the limits of our analytical model versus the medium birefringence and maximum rotation angle, both for weakly and highly anisotropic materials.  \\
Our findings pave the way to the design and realization of a new kind of electromagnetic waveguides, with the presented mechanism valid regardless of frequency. Twisted structures can be realized, for example, using liquid crystals \cite{Marrucci:2006,Dalessandro:2015,Kobashi:2016}, laser-written anisotropic glasses \cite{Ling:2015}, metamaterials \cite{Bomzon:2002,Lin:2014,Askarpour:2014}. With respect to Ref.~\cite{Slussarenko:2016}, the geometry  proposed hereby is much simpler to realize as it is structured only in the transverse plane. Furthermore, the longitudinal homogeneity of the sample inhibits back reflections, an important detrimental effect when fabricating high-quality waveguides. Our results can be generalized to more complicated and exotic anisotropic responses, including, e.g., magneto-electric coupling \cite{Pendry:1999,Smith:2003,Ghosh:2008,Khanikaev:2013,Pfeiffer:2013,Bliokh:2014,Asadchy:2015}. Several future developments can be envisaged, including the systematic investigation of light trapping versus geometrical and material parameters of the structure, and the usage of our model to find the periodic components of the quasi-mode. It will be also interesting to investigate the interplay between the Rytov-Vladimirskii-Berry and the Pancharatnam-Berry phases which are both of geometric origin \cite{Bliokh:2015}, and the connection of our system with the recently introduced gauge optics  \cite{Fang:2013,Lin:2014_1}. \\ 
From a more fundamental point of view, this works is an important contribution to the analogy between light propagation in anisotropic materials and propagation of spin 1/2 particles in magnetic fields: in fact, confinement based on geometric phase can be readily transposed to matter waves, e.g., charged particles in a Paul trap \cite{Paul:1990}. 

\section*{Acknowledgments} C.P.J. gratefully acknowledges Funda\c{c}\~{a}o para a Ci\^{e}ncia e a Tecnologia, POPH-QREN and FSE (FCT, Portugal) for the fellowship SFRH/BPD/77524/2011; she also thanks the Optics Laboratory in Tampere for the hospitality. A.A. and G.A. thank the Academy of Finland through the Finland Distinguished Professor grant no. 282858. L.M. acknowledges support from the European Research Council (ERC), under grant No. 694683, PHOSPhOR.


\begin{thebibliography}{66}%
\makeatletter
\providecommand \@ifxundefined [1]{%
 \@ifx{#1\undefined}
}%
\providecommand \@ifnum [1]{%
 \ifnum #1\expandafter \@firstoftwo
 \else \expandafter \@secondoftwo
 \fi
}%
\providecommand \@ifx [1]{%
 \ifx #1\expandafter \@firstoftwo
 \else \expandafter \@secondoftwo
 \fi
}%
\providecommand \natexlab [1]{#1}%
\providecommand \enquote  [1]{``#1''}%
\providecommand \bibnamefont  [1]{#1}%
\providecommand \bibfnamefont [1]{#1}%
\providecommand \citenamefont [1]{#1}%
\providecommand \href@noop [0]{\@secondoftwo}%
\providecommand \href [0]{\begingroup \@sanitize@url \@href}%
\providecommand \@href[1]{\@@startlink{#1}\@@href}%
\providecommand \@@href[1]{\endgroup#1\@@endlink}%
\providecommand \@sanitize@url [0]{\catcode `\\12\catcode `\$12\catcode
  `\&12\catcode `\#12\catcode `\^12\catcode `\_12\catcode `\%12\relax}%
\providecommand \@@startlink[1]{}%
\providecommand \@@endlink[0]{}%
\providecommand \url  [0]{\begingroup\@sanitize@url \@url }%
\providecommand \@url [1]{\endgroup\@href {#1}{\urlprefix }}%
\providecommand \urlprefix  [0]{URL }%
\providecommand \Eprint [0]{\href }%
\providecommand \doibase [0]{http://dx.doi.org/}%
\providecommand \selectlanguage [0]{\@gobble}%
\providecommand \bibinfo  [0]{\@secondoftwo}%
\providecommand \bibfield  [0]{\@secondoftwo}%
\providecommand \translation [1]{[#1]}%
\providecommand \BibitemOpen [0]{}%
\providecommand \bibitemStop [0]{}%
\providecommand \bibitemNoStop [0]{.\EOS\space}%
\providecommand \EOS [0]{\spacefactor3000\relax}%
\providecommand \BibitemShut  [1]{\csname bibitem#1\endcsname}%
\let\auto@bib@innerbib\@empty
\bibitem [{\citenamefont {Kildishev}\ \emph {et~al.}(2013)\citenamefont
  {Kildishev}, \citenamefont {Boltasseva},\ and\ \citenamefont
  {Shalaev}}]{Kildishev:2013}%
  \BibitemOpen
  \bibfield  {author} {\bibinfo {author} {\bibfnamefont {Alexander~V.}\
  \bibnamefont {Kildishev}}, \bibinfo {author} {\bibfnamefont {Alexandra}\
  \bibnamefont {Boltasseva}}, \ and\ \bibinfo {author} {\bibfnamefont
  {Vladimir~M.}\ \bibnamefont {Shalaev}},\ }\bibfield  {title} {\enquote
  {\bibinfo {title} {Planar photonics with metasurfaces},}\ }\href {\doibase
  10.1126/science.1232009} {\bibfield  {journal} {\bibinfo  {journal}
  {Science}\ }\textbf {\bibinfo {volume} {339}} (\bibinfo {year} {2013}),\
  10.1126/science.1232009},\ \Eprint
  {http://arxiv.org/abs/http://science.sciencemag.org/content/339/6125/1232009.full.pdf}
  {http://science.sciencemag.org/content/339/6125/1232009.full.pdf}
  \BibitemShut {NoStop}%
\bibitem [{\citenamefont {Yu}\ and\ \citenamefont {Capasso}(2014)}]{Yu:2014}%
  \BibitemOpen
  \bibfield  {author} {\bibinfo {author} {\bibfnamefont {Nanfang}\ \bibnamefont
  {Yu}}\ and\ \bibinfo {author} {\bibfnamefont {Federico}\ \bibnamefont
  {Capasso}},\ }\bibfield  {title} {\enquote {\bibinfo {title} {Flat optics
  with designer metasurfaces},}\ }\href@noop {} {\bibfield  {journal} {\bibinfo
   {journal} {Nat. Mater.}\ }\textbf {\bibinfo {volume} {13}},\ \bibinfo
  {pages} {139--150} (\bibinfo {year} {2014})}\BibitemShut {NoStop}%
\bibitem [{\citenamefont {Lin}\ \emph {et~al.}(2014)\citenamefont {Lin},
  \citenamefont {Fan}, \citenamefont {Hasman},\ and\ \citenamefont
  {Brongersma}}]{Lin:2014}%
  \BibitemOpen
  \bibfield  {author} {\bibinfo {author} {\bibfnamefont {Dianmin}\ \bibnamefont
  {Lin}}, \bibinfo {author} {\bibfnamefont {Pengyu}\ \bibnamefont {Fan}},
  \bibinfo {author} {\bibfnamefont {Erez}\ \bibnamefont {Hasman}}, \ and\
  \bibinfo {author} {\bibfnamefont {Mark~L.}\ \bibnamefont {Brongersma}},\
  }\bibfield  {title} {\enquote {\bibinfo {title} {Dielectric gradient
  metasurface optical elements},}\ }\href {\doibase 10.1126/science.1253213}
  {\bibfield  {journal} {\bibinfo  {journal} {Science}\ }\textbf {\bibinfo
  {volume} {345}},\ \bibinfo {pages} {298--302} (\bibinfo {year}
  {2014})}\BibitemShut {NoStop}%
\bibitem [{\citenamefont {Kobashi}\ \emph
  {et~al.}(2016{\natexlab{a}})\citenamefont {Kobashi}, \citenamefont
  {Yoshida},\ and\ \citenamefont {Ozaki}}]{Kobashi:2016}%
  \BibitemOpen
  \bibfield  {author} {\bibinfo {author} {\bibfnamefont {Junji}\ \bibnamefont
  {Kobashi}}, \bibinfo {author} {\bibfnamefont {Hiroyuki}\ \bibnamefont
  {Yoshida}}, \ and\ \bibinfo {author} {\bibfnamefont {Masanori}\ \bibnamefont
  {Ozaki}},\ }\bibfield  {title} {\enquote {\bibinfo {title} {Planar optics
  with patterned chiral liquid crystals},}\ }\href@noop {} {\bibfield
  {journal} {\bibinfo  {journal} {Nat. Photon.}\ }\textbf {\bibinfo {volume}
  {10}},\ \bibinfo {pages} {389--392} (\bibinfo {year}
  {2016}{\natexlab{a}})}\BibitemShut {NoStop}%
\bibitem [{\citenamefont {Bomzon}\ \emph {et~al.}(2002)\citenamefont {Bomzon},
  \citenamefont {Biener}, \citenamefont {Kleiner},\ and\ \citenamefont
  {Hasman}}]{Bomzon:2002}%
  \BibitemOpen
  \bibfield  {author} {\bibinfo {author} {\bibfnamefont {Ze'ev}\ \bibnamefont
  {Bomzon}}, \bibinfo {author} {\bibfnamefont {Gabriel}\ \bibnamefont
  {Biener}}, \bibinfo {author} {\bibfnamefont {Vladimir}\ \bibnamefont
  {Kleiner}}, \ and\ \bibinfo {author} {\bibfnamefont {Erez}\ \bibnamefont
  {Hasman}},\ }\bibfield  {title} {\enquote {\bibinfo {title} {Space-variant
  {P}ancharatnam--{B}erry phase optical elements with computer-generated
  subwavelength gratings},}\ }\href {\doibase 10.1364/OL.27.001141} {\bibfield
  {journal} {\bibinfo  {journal} {Opt. Lett.}\ }\textbf {\bibinfo {volume}
  {27}},\ \bibinfo {pages} {1141--1143} (\bibinfo {year} {2002})}\BibitemShut
  {NoStop}%
\bibitem [{\citenamefont {Marrucci}\ \emph
  {et~al.}(2006{\natexlab{a}})\citenamefont {Marrucci}, \citenamefont {Manzo},\
  and\ \citenamefont {Paparo}}]{Marrucci:2006_1}%
  \BibitemOpen
  \bibfield  {author} {\bibinfo {author} {\bibfnamefont {L.}~\bibnamefont
  {Marrucci}}, \bibinfo {author} {\bibfnamefont {C.}~\bibnamefont {Manzo}}, \
  and\ \bibinfo {author} {\bibfnamefont {D.}~\bibnamefont {Paparo}},\
  }\bibfield  {title} {\enquote {\bibinfo {title} {Pancharatnam--{B}erry phase
  optical elements for wavefront shaping in the visible domain: switchable
  helical modes generation},}\ }\href@noop {} {\bibfield  {journal} {\bibinfo
  {journal} {Appl. Phys. Lett.}\ }\textbf {\bibinfo {volume} {88}},\ \bibinfo
  {pages} {221102} (\bibinfo {year} {2006}{\natexlab{a}})}\BibitemShut
  {NoStop}%
\bibitem [{\citenamefont {Berry}(1987)}]{Berry:1987}%
  \BibitemOpen
  \bibfield  {author} {\bibinfo {author} {\bibfnamefont {M.V.}\ \bibnamefont
  {Berry}},\ }\bibfield  {title} {\enquote {\bibinfo {title} {The adiabatic
  phase and {P}ancharatnam's phase for polarized light},}\ }\href@noop {}
  {\bibfield  {journal} {\bibinfo  {journal} {J. Mod. Opt.}\ }\textbf {\bibinfo
  {volume} {34}},\ \bibinfo {pages} {1401--1407} (\bibinfo {year}
  {1987})}\BibitemShut {NoStop}%
\bibitem [{\citenamefont {Bomzon}\ \emph {et~al.}(2001)\citenamefont {Bomzon},
  \citenamefont {Kleiner},\ and\ \citenamefont {Hasman}}]{Bomzon:2001}%
  \BibitemOpen
  \bibfield  {author} {\bibinfo {author} {\bibfnamefont {Ze’ev}\ \bibnamefont
  {Bomzon}}, \bibinfo {author} {\bibfnamefont {Vladimir}\ \bibnamefont
  {Kleiner}}, \ and\ \bibinfo {author} {\bibfnamefont {Erez}\ \bibnamefont
  {Hasman}},\ }\bibfield  {title} {\enquote {\bibinfo {title} {Formation of
  radially and azimuthally polarized light using space-variant subwavelength
  metal stripe gratings},}\ }\href {\doibase
  http://dx.doi.org/10.1063/1.1401091} {\bibfield  {journal} {\bibinfo
  {journal} {Appl. Phys. Lett.}\ }\textbf {\bibinfo {volume} {79}},\ \bibinfo
  {pages} {1587--1589} (\bibinfo {year} {2001})}\BibitemShut {NoStop}%
\bibitem [{\citenamefont {Hasman}\ \emph {et~al.}(2003)\citenamefont {Hasman},
  \citenamefont {Kleiner}, \citenamefont {Biener},\ and\ \citenamefont
  {Niv}}]{Hasman:2003}%
  \BibitemOpen
  \bibfield  {author} {\bibinfo {author} {\bibfnamefont {Erez}\ \bibnamefont
  {Hasman}}, \bibinfo {author} {\bibfnamefont {Vladimir}\ \bibnamefont
  {Kleiner}}, \bibinfo {author} {\bibfnamefont {Gabriel}\ \bibnamefont
  {Biener}}, \ and\ \bibinfo {author} {\bibfnamefont {Avi}\ \bibnamefont
  {Niv}},\ }\bibfield  {title} {\enquote {\bibinfo {title} {Polarization
  dependent focusing lens by use of quantized pancharatnam–berry phase
  diffractive optics},}\ }\href {\doibase http://dx.doi.org/10.1063/1.1539300}
  {\bibfield  {journal} {\bibinfo  {journal} {Appl. Phys. Lett.}\ }\textbf
  {\bibinfo {volume} {82}},\ \bibinfo {pages} {328--330} (\bibinfo {year}
  {2003})}\BibitemShut {NoStop}%
\bibitem [{\citenamefont {Marrucci}\ \emph
  {et~al.}(2006{\natexlab{b}})\citenamefont {Marrucci}, \citenamefont {Manzo},\
  and\ \citenamefont {Paparo}}]{Marrucci:2006}%
  \BibitemOpen
  \bibfield  {author} {\bibinfo {author} {\bibfnamefont {L.}~\bibnamefont
  {Marrucci}}, \bibinfo {author} {\bibfnamefont {C.}~\bibnamefont {Manzo}}, \
  and\ \bibinfo {author} {\bibfnamefont {D.}~\bibnamefont {Paparo}},\
  }\bibfield  {title} {\enquote {\bibinfo {title} {Optical spin-to-orbital
  angular momentum conversion in inhomogeneous anisotropic media},}\
  }\href@noop {} {\bibfield  {journal} {\bibinfo  {journal} {Phys. Rev. Lett.}\
  }\textbf {\bibinfo {volume} {96}},\ \bibinfo {pages} {163905} (\bibinfo
  {year} {2006}{\natexlab{b}})}\BibitemShut {NoStop}%
\bibitem [{\citenamefont {Slussarenko}\ \emph {et~al.}(2011)\citenamefont
  {Slussarenko}, \citenamefont {Murauski}, \citenamefont {Du}, \citenamefont
  {Chigrinov}, \citenamefont {Marrucci},\ and\ \citenamefont
  {Santamato}}]{Slussarenko:2011}%
  \BibitemOpen
  \bibfield  {author} {\bibinfo {author} {\bibfnamefont {S.}~\bibnamefont
  {Slussarenko}}, \bibinfo {author} {\bibfnamefont {A.}~\bibnamefont
  {Murauski}}, \bibinfo {author} {\bibfnamefont {T.}~\bibnamefont {Du}},
  \bibinfo {author} {\bibfnamefont {V.}~\bibnamefont {Chigrinov}}, \bibinfo
  {author} {\bibfnamefont {L.}~\bibnamefont {Marrucci}}, \ and\ \bibinfo
  {author} {\bibfnamefont {E.}~\bibnamefont {Santamato}},\ }\bibfield  {title}
  {\enquote {\bibinfo {title} {Tunable liquid crystal q-plates with arbitrary
  topological charge},}\ }\href@noop {} {\bibfield  {journal} {\bibinfo
  {journal} {Opt. Express}\ }\textbf {\bibinfo {volume} {19}},\ \bibinfo
  {pages} {4085--4090} (\bibinfo {year} {2011})}\BibitemShut {NoStop}%
\bibitem [{\citenamefont {Kim}\ \emph {et~al.}(2015)\citenamefont {Kim},
  \citenamefont {Li}, \citenamefont {Miskiewicz}, \citenamefont {Oh},
  \citenamefont {Kudenov},\ and\ \citenamefont {Escuti}}]{Kim:2015}%
  \BibitemOpen
  \bibfield  {author} {\bibinfo {author} {\bibfnamefont {Jihwan}\ \bibnamefont
  {Kim}}, \bibinfo {author} {\bibfnamefont {Yanming}\ \bibnamefont {Li}},
  \bibinfo {author} {\bibfnamefont {Matthew~N.}\ \bibnamefont {Miskiewicz}},
  \bibinfo {author} {\bibfnamefont {Chulwoo}\ \bibnamefont {Oh}}, \bibinfo
  {author} {\bibfnamefont {Michael~W.}\ \bibnamefont {Kudenov}}, \ and\
  \bibinfo {author} {\bibfnamefont {Michael~J.}\ \bibnamefont {Escuti}},\
  }\bibfield  {title} {\enquote {\bibinfo {title} {Fabrication of ideal
  geometric-phase holograms with arbitrary wavefronts},}\ }\href {\doibase
  10.1364/OPTICA.2.000958} {\bibfield  {journal} {\bibinfo  {journal} {Optica}\
  }\textbf {\bibinfo {volume} {2}},\ \bibinfo {pages} {958--964} (\bibinfo
  {year} {2015})}\BibitemShut {NoStop}%
\bibitem [{\citenamefont {Bauer}\ \emph {et~al.}(2015)\citenamefont {Bauer},
  \citenamefont {Banzer}, \citenamefont {Karimi}, \citenamefont {Orlov},
  \citenamefont {Rubano}, \citenamefont {Marrucci}, \citenamefont {Santamato},
  \citenamefont {Boyd},\ and\ \citenamefont {Leuchs}}]{Bauer:2015}%
  \BibitemOpen
  \bibfield  {author} {\bibinfo {author} {\bibfnamefont {Thomas}\ \bibnamefont
  {Bauer}}, \bibinfo {author} {\bibfnamefont {Peter}\ \bibnamefont {Banzer}},
  \bibinfo {author} {\bibfnamefont {Ebrahim}\ \bibnamefont {Karimi}}, \bibinfo
  {author} {\bibfnamefont {Sergej}\ \bibnamefont {Orlov}}, \bibinfo {author}
  {\bibfnamefont {Andrea}\ \bibnamefont {Rubano}}, \bibinfo {author}
  {\bibfnamefont {Lorenzo}\ \bibnamefont {Marrucci}}, \bibinfo {author}
  {\bibfnamefont {Enrico}\ \bibnamefont {Santamato}}, \bibinfo {author}
  {\bibfnamefont {Robert~W.}\ \bibnamefont {Boyd}}, \ and\ \bibinfo {author}
  {\bibfnamefont {Gerd}\ \bibnamefont {Leuchs}},\ }\bibfield  {title} {\enquote
  {\bibinfo {title} {Observation of optical polarization {M}\"obius strips},}\
  }\href {\doibase 10.1126/science.1260635} {\bibfield  {journal} {\bibinfo
  {journal} {Science}\ }\textbf {\bibinfo {volume} {347}},\ \bibinfo {pages}
  {964--966} (\bibinfo {year} {2015})},\ \Eprint
  {http://arxiv.org/abs/http://www.sciencemag.org/content/347/6225/964.full.pdf}
  {http://www.sciencemag.org/content/347/6225/964.full.pdf} \BibitemShut
  {NoStop}%
\bibitem [{\citenamefont {Naidoo}\ \emph {et~al.}(2016)\citenamefont {Naidoo},
  \citenamefont {Roux}, \citenamefont {Dudley}, \citenamefont {Litvin},
  \citenamefont {Piccirillo}, \citenamefont {Marrucci},\ and\ \citenamefont
  {Forbes}}]{Naidoo:2016}%
  \BibitemOpen
  \bibfield  {author} {\bibinfo {author} {\bibfnamefont {Darryl}\ \bibnamefont
  {Naidoo}}, \bibinfo {author} {\bibfnamefont {Filippus~S.}\ \bibnamefont
  {Roux}}, \bibinfo {author} {\bibfnamefont {Angela}\ \bibnamefont {Dudley}},
  \bibinfo {author} {\bibfnamefont {Igor}\ \bibnamefont {Litvin}}, \bibinfo
  {author} {\bibfnamefont {Bruno}\ \bibnamefont {Piccirillo}}, \bibinfo
  {author} {\bibfnamefont {Lorenzo}\ \bibnamefont {Marrucci}}, \ and\ \bibinfo
  {author} {\bibfnamefont {Andrew}\ \bibnamefont {Forbes}},\ }\bibfield
  {title} {\enquote {\bibinfo {title} {Controlled generation of higher-order
  poincaré sphere beams from a laser},}\ }\href@noop {} {\bibfield  {journal}
  {\bibinfo  {journal} {Nat. Photon.}\ }\textbf {\bibinfo {volume} {10}},\
  \bibinfo {pages} {327--332} (\bibinfo {year} {2016})}\BibitemShut {NoStop}%
\bibitem [{\citenamefont {Todorov}\ and\ \citenamefont
  {Nikolova}(1992)}]{Todorov:1992}%
  \BibitemOpen
  \bibfield  {author} {\bibinfo {author} {\bibfnamefont {T.}~\bibnamefont
  {Todorov}}\ and\ \bibinfo {author} {\bibfnamefont {L.}~\bibnamefont
  {Nikolova}},\ }\bibfield  {title} {\enquote {\bibinfo {title}
  {Spectrophotopolarimeter: fast simultaneous real-time measurement of light
  parameters},}\ }\href@noop {} {\bibfield  {journal} {\bibinfo  {journal}
  {Opt. Lett.}\ }\textbf {\bibinfo {volume} {17}},\ \bibinfo {pages} {358--359}
  (\bibinfo {year} {1992})}\BibitemShut {NoStop}%
\bibitem [{\citenamefont {Gori}(1999)}]{Gori:1999}%
  \BibitemOpen
  \bibfield  {author} {\bibinfo {author} {\bibfnamefont {Franco}\ \bibnamefont
  {Gori}},\ }\bibfield  {title} {\enquote {\bibinfo {title} {Measuring stokes
  parameters by means of a polarization grating},}\ }\href {\doibase
  10.1364/OL.24.000584} {\bibfield  {journal} {\bibinfo  {journal} {Opt.
  Lett.}\ }\textbf {\bibinfo {volume} {24}},\ \bibinfo {pages} {584--586}
  (\bibinfo {year} {1999})}\BibitemShut {NoStop}%
\bibitem [{\citenamefont {Provenzano}\ \emph {et~al.}(2007)\citenamefont
  {Provenzano}, \citenamefont {Pagliusi},\ and\ \citenamefont
  {Cipparrone}}]{Provenzano:2007}%
  \BibitemOpen
  \bibfield  {author} {\bibinfo {author} {\bibfnamefont {C.}~\bibnamefont
  {Provenzano}}, \bibinfo {author} {\bibfnamefont {P.}~\bibnamefont
  {Pagliusi}}, \ and\ \bibinfo {author} {\bibfnamefont {G.}~\bibnamefont
  {Cipparrone}},\ }\bibfield  {title} {\enquote {\bibinfo {title} {Electrically
  tunable two-dimensional liquid crystals gratings induced by polarization
  holography},}\ }\href {\doibase 10.1364/OE.15.005872} {\bibfield  {journal}
  {\bibinfo  {journal} {Opt. Express}\ }\textbf {\bibinfo {volume} {15}},\
  \bibinfo {pages} {5872--5878} (\bibinfo {year} {2007})}\BibitemShut {NoStop}%
\bibitem [{\citenamefont {Tabiryan}\ \emph {et~al.}(2011)\citenamefont
  {Tabiryan}, \citenamefont {Nersisyan}, \citenamefont {White}, \citenamefont
  {Bunning}, \citenamefont {Steeves},\ and\ \citenamefont
  {Kimball}}]{Tabiryan:2011}%
  \BibitemOpen
  \bibfield  {author} {\bibinfo {author} {\bibfnamefont {Nelson~V.}\
  \bibnamefont {Tabiryan}}, \bibinfo {author} {\bibfnamefont {Sarik~R.}\
  \bibnamefont {Nersisyan}}, \bibinfo {author} {\bibfnamefont {Timothy~J.}\
  \bibnamefont {White}}, \bibinfo {author} {\bibfnamefont {Timothy~J.}\
  \bibnamefont {Bunning}}, \bibinfo {author} {\bibfnamefont {Diane~M.}\
  \bibnamefont {Steeves}}, \ and\ \bibinfo {author} {\bibfnamefont {Brian~R.}\
  \bibnamefont {Kimball}},\ }\bibfield  {title} {\enquote {\bibinfo {title}
  {Transparent thin film polarizing and optical control systems},}\ }\href
  {\doibase http://dx.doi.org/10.1063/1.3609965} {\bibfield  {journal}
  {\bibinfo  {journal} {AIP Advances}\ }\textbf {\bibinfo {volume} {1}},\
  \bibinfo {eid} {022153} (\bibinfo {year} {2011}),\
  http://dx.doi.org/10.1063/1.3609965}\BibitemShut {NoStop}%
\bibitem [{\citenamefont {Ruiz}\ \emph {et~al.}(2013)\citenamefont {Ruiz},
  \citenamefont {Pagliusi}, \citenamefont {Provenzano},\ and\ \citenamefont
  {Cipparrone}}]{Ruiz:2013}%
  \BibitemOpen
  \bibfield  {author} {\bibinfo {author} {\bibfnamefont {U.}~\bibnamefont
  {Ruiz}}, \bibinfo {author} {\bibfnamefont {P.}~\bibnamefont {Pagliusi}},
  \bibinfo {author} {\bibfnamefont {C.}~\bibnamefont {Provenzano}}, \ and\
  \bibinfo {author} {\bibfnamefont {G.}~\bibnamefont {Cipparrone}},\ }\bibfield
   {title} {\enquote {\bibinfo {title} {Highly efficient generation of vector
  beams through polarization holograms},}\ }\href {\doibase
  http://dx.doi.org/10.1063/1.4801317} {\bibfield  {journal} {\bibinfo
  {journal} {Appl. Phys. Lett.}\ }\textbf {\bibinfo {volume} {102}},\ \bibinfo
  {eid} {161104} (\bibinfo {year} {2013}),\
  http://dx.doi.org/10.1063/1.4801317}\BibitemShut {NoStop}%
\bibitem [{\citenamefont {Litchinitser}(2016)}]{Litchinitser:2016}%
  \BibitemOpen
  \bibfield  {author} {\bibinfo {author} {\bibfnamefont {Natalia~M.}\
  \bibnamefont {Litchinitser}},\ }\bibfield  {title} {\enquote {\bibinfo
  {title} {Photonic multitasking enabled with geometric phase},}\ }\href
  {\doibase 10.1126/science.aaf8391} {\bibfield  {journal} {\bibinfo  {journal}
  {Science}\ }\textbf {\bibinfo {volume} {352}},\ \bibinfo {pages} {1177--1178}
  (\bibinfo {year} {2016})},\ \Eprint
  {http://arxiv.org/abs/http://science.sciencemag.org/content/352/6290/1177.full.pdf}
  {http://science.sciencemag.org/content/352/6290/1177.full.pdf} \BibitemShut
  {NoStop}%
\bibitem [{\citenamefont {Roux}(2006)}]{Roux:2006}%
  \BibitemOpen
  \bibfield  {author} {\bibinfo {author} {\bibfnamefont {Filippus~S.}\
  \bibnamefont {Roux}},\ }\bibfield  {title} {\enquote {\bibinfo {title}
  {Geometric phase lens},}\ }\href@noop {} {\bibfield  {journal} {\bibinfo
  {journal} {J. Opt. Soc. Am. B}\ }\textbf {\bibinfo {volume} {23}},\ \bibinfo
  {pages} {476--482} (\bibinfo {year} {2006})}\BibitemShut {NoStop}%
\bibitem [{\citenamefont {Khorasaninejad}\ \emph {et~al.}(2016)\citenamefont
  {Khorasaninejad}, \citenamefont {Chen}, \citenamefont {Devlin}, \citenamefont
  {Oh}, \citenamefont {Zhu},\ and\ \citenamefont
  {Capasso}}]{Khorasaninejad:2016}%
  \BibitemOpen
  \bibfield  {author} {\bibinfo {author} {\bibfnamefont {Mohammadreza}\
  \bibnamefont {Khorasaninejad}}, \bibinfo {author} {\bibfnamefont {Wei~Ting}\
  \bibnamefont {Chen}}, \bibinfo {author} {\bibfnamefont {Robert~C.}\
  \bibnamefont {Devlin}}, \bibinfo {author} {\bibfnamefont {Jaewon}\
  \bibnamefont {Oh}}, \bibinfo {author} {\bibfnamefont {Alexander~Y.}\
  \bibnamefont {Zhu}}, \ and\ \bibinfo {author} {\bibfnamefont {Federico}\
  \bibnamefont {Capasso}},\ }\bibfield  {title} {\enquote {\bibinfo {title}
  {Metalenses at visible wavelengths: Diffraction-limited focusing and
  subwavelength resolution imaging},}\ }\href {\doibase
  10.1126/science.aaf6644} {\bibfield  {journal} {\bibinfo  {journal}
  {Science}\ }\textbf {\bibinfo {volume} {352}},\ \bibinfo {pages} {1190--1194}
  (\bibinfo {year} {2016})},\ \Eprint
  {http://arxiv.org/abs/http://science.sciencemag.org/content/352/6290/1190.full.pdf}
  {http://science.sciencemag.org/content/352/6290/1190.full.pdf} \BibitemShut
  {NoStop}%
\bibitem [{\citenamefont {Zheng}\ \emph {et~al.}(2015)\citenamefont {Zheng},
  \citenamefont {Mühlenbernd}, \citenamefont {Kenney}, \citenamefont {Li},
  \citenamefont {Zentgraf},\ and\ \citenamefont {Zhang}}]{Zheng:2015}%
  \BibitemOpen
  \bibfield  {author} {\bibinfo {author} {\bibfnamefont {Guoxing}\ \bibnamefont
  {Zheng}}, \bibinfo {author} {\bibfnamefont {Holger}\ \bibnamefont
  {Mühlenbernd}}, \bibinfo {author} {\bibfnamefont {Mitchell}\ \bibnamefont
  {Kenney}}, \bibinfo {author} {\bibfnamefont {Guixin}\ \bibnamefont {Li}},
  \bibinfo {author} {\bibfnamefont {Thomas}\ \bibnamefont {Zentgraf}}, \ and\
  \bibinfo {author} {\bibfnamefont {Shuang}\ \bibnamefont {Zhang}},\ }\bibfield
   {title} {\enquote {\bibinfo {title} {Metasurface holograms reaching 80$\%$
  efficiency},}\ }\href@noop {} {\bibfield  {journal} {\bibinfo  {journal}
  {Nat. Nanotech.}\ }\textbf {\bibinfo {volume} {10}},\ \bibinfo {pages}
  {308--312} (\bibinfo {year} {2015})}\BibitemShut {NoStop}%
\bibitem [{\citenamefont {Nersisyan}\ \emph {et~al.}(2009)\citenamefont
  {Nersisyan}, \citenamefont {Tabiryan}, \citenamefont {Steeves},\ and\
  \citenamefont {Kimball}}]{Nersisyan:2009}%
  \BibitemOpen
  \bibfield  {author} {\bibinfo {author} {\bibfnamefont {Sarik}\ \bibnamefont
  {Nersisyan}}, \bibinfo {author} {\bibfnamefont {Nelson}\ \bibnamefont
  {Tabiryan}}, \bibinfo {author} {\bibfnamefont {Diane~M.}\ \bibnamefont
  {Steeves}}, \ and\ \bibinfo {author} {\bibfnamefont {Brian~R.}\ \bibnamefont
  {Kimball}},\ }\bibfield  {title} {\enquote {\bibinfo {title} {Fabrication of
  liquid crystal polymer axial waveplates for uv-ir wavelengths},}\ }\href
  {\doibase 10.1364/OE.17.011926} {\bibfield  {journal} {\bibinfo  {journal}
  {Opt. Express}\ }\textbf {\bibinfo {volume} {17}},\ \bibinfo {pages}
  {11926--11934} (\bibinfo {year} {2009})}\BibitemShut {NoStop}%
\bibitem [{\citenamefont {Li}\ \emph {et~al.}(2013)\citenamefont {Li},
  \citenamefont {Kang}, \citenamefont {Chen}, \citenamefont {Zhang},
  \citenamefont {Pun}, \citenamefont {Cheah},\ and\ \citenamefont
  {Li}}]{Li:2013}%
  \BibitemOpen
  \bibfield  {author} {\bibinfo {author} {\bibfnamefont {Guixin}\ \bibnamefont
  {Li}}, \bibinfo {author} {\bibfnamefont {Ming}\ \bibnamefont {Kang}},
  \bibinfo {author} {\bibfnamefont {Shumei}\ \bibnamefont {Chen}}, \bibinfo
  {author} {\bibfnamefont {Shuang}\ \bibnamefont {Zhang}}, \bibinfo {author}
  {\bibfnamefont {Edwin Yue-Bun}\ \bibnamefont {Pun}}, \bibinfo {author}
  {\bibfnamefont {K.~W.}\ \bibnamefont {Cheah}}, \ and\ \bibinfo {author}
  {\bibfnamefont {Jensen}\ \bibnamefont {Li}},\ }\bibfield  {title} {\enquote
  {\bibinfo {title} {Spin-enabled plasmonic metasurfaces for manipulating
  orbital angular momentum of light},}\ }\href {\doibase 10.1021/nl401734r}
  {\bibfield  {journal} {\bibinfo  {journal} {Nano Lett.}\ }\textbf {\bibinfo
  {volume} {13}},\ \bibinfo {pages} {4148--4151} (\bibinfo {year} {2013})},\
  \bibinfo {note} {pMID: 23965168},\ \Eprint
  {http://arxiv.org/abs/http://dx.doi.org/10.1021/nl401734r}
  {http://dx.doi.org/10.1021/nl401734r} \BibitemShut {NoStop}%
\bibitem [{\citenamefont {Ling}\ \emph {et~al.}(2015)\citenamefont {Ling},
  \citenamefont {Zhou}, \citenamefont {Yi}, \citenamefont {Shu}, \citenamefont
  {Liu}, \citenamefont {Chen}, \citenamefont {Luo}, \citenamefont {Wen},\ and\
  \citenamefont {Fan}}]{Ling:2015}%
  \BibitemOpen
  \bibfield  {author} {\bibinfo {author} {\bibfnamefont {Xiaohui}\ \bibnamefont
  {Ling}}, \bibinfo {author} {\bibfnamefont {Xinxing}\ \bibnamefont {Zhou}},
  \bibinfo {author} {\bibfnamefont {Xunong}\ \bibnamefont {Yi}}, \bibinfo
  {author} {\bibfnamefont {Weixing}\ \bibnamefont {Shu}}, \bibinfo {author}
  {\bibfnamefont {Yachao}\ \bibnamefont {Liu}}, \bibinfo {author}
  {\bibfnamefont {Shizhen}\ \bibnamefont {Chen}}, \bibinfo {author}
  {\bibfnamefont {Hailu}\ \bibnamefont {Luo}}, \bibinfo {author} {\bibfnamefont
  {Shuangchun}\ \bibnamefont {Wen}}, \ and\ \bibinfo {author} {\bibfnamefont
  {Dianyuan}\ \bibnamefont {Fan}},\ }\bibfield  {title} {\enquote {\bibinfo
  {title} {Giant photonic spin hall effect in momentum space in a structured
  metamaterial with spatially varying birefringence},}\ }\href@noop {}
  {\bibfield  {journal} {\bibinfo  {journal} {Light Sci. Appl.}\ }\textbf
  {\bibinfo {volume} {4}},\ \bibinfo {pages} {e290} (\bibinfo {year}
  {2015})}\BibitemShut {NoStop}%
\bibitem [{\citenamefont {Rafayelyan}\ \emph {et~al.}(2016)\citenamefont
  {Rafayelyan}, \citenamefont {Tkachenko},\ and\ \citenamefont
  {Brasselet}}]{Rafayelyan:2016}%
  \BibitemOpen
  \bibfield  {author} {\bibinfo {author} {\bibfnamefont {Mushegh}\ \bibnamefont
  {Rafayelyan}}, \bibinfo {author} {\bibfnamefont {Georgiy}\ \bibnamefont
  {Tkachenko}}, \ and\ \bibinfo {author} {\bibfnamefont {Etienne}\ \bibnamefont
  {Brasselet}},\ }\bibfield  {title} {\enquote {\bibinfo {title} {Reflective
  spin-orbit geometric phase from chiral anisotropic optical media},}\ }\href
  {\doibase 10.1103/PhysRevLett.116.253902} {\bibfield  {journal} {\bibinfo
  {journal} {Phys. Rev. Lett.}\ }\textbf {\bibinfo {volume} {116}},\ \bibinfo
  {pages} {253902} (\bibinfo {year} {2016})}\BibitemShut {NoStop}%
\bibitem [{\citenamefont {Kobashi}\ \emph
  {et~al.}(2016{\natexlab{b}})\citenamefont {Kobashi}, \citenamefont
  {Yoshida},\ and\ \citenamefont {Ozaki}}]{Kobashi:2016_1}%
  \BibitemOpen
  \bibfield  {author} {\bibinfo {author} {\bibfnamefont {Junji}\ \bibnamefont
  {Kobashi}}, \bibinfo {author} {\bibfnamefont {Hiroyuki}\ \bibnamefont
  {Yoshida}}, \ and\ \bibinfo {author} {\bibfnamefont {Masanori}\ \bibnamefont
  {Ozaki}},\ }\bibfield  {title} {\enquote {\bibinfo {title} {Polychromatic
  optical vortex generation from patterned cholesteric liquid crystals},}\
  }\href {\doibase 10.1103/PhysRevLett.116.253903} {\bibfield  {journal}
  {\bibinfo  {journal} {Phys. Rev. Lett.}\ }\textbf {\bibinfo {volume} {116}},\
  \bibinfo {pages} {253903} (\bibinfo {year} {2016}{\natexlab{b}})}\BibitemShut
  {NoStop}%
\bibitem [{\citenamefont {Barboza}\ \emph {et~al.}(2016)\citenamefont
  {Barboza}, \citenamefont {Bortolozzo}, \citenamefont {Residori},\ and\
  \citenamefont {Clerc}}]{Barboza:2016}%
  \BibitemOpen
  \bibfield  {author} {\bibinfo {author} {\bibfnamefont {Raouf}\ \bibnamefont
  {Barboza}}, \bibinfo {author} {\bibfnamefont {Umberto}\ \bibnamefont
  {Bortolozzo}}, \bibinfo {author} {\bibfnamefont {Stefania}\ \bibnamefont
  {Residori}}, \ and\ \bibinfo {author} {\bibfnamefont {Marcel~G.}\
  \bibnamefont {Clerc}},\ }\bibfield  {title} {\enquote {\bibinfo {title}
  {Berry phase of light bragg-reflected by chiral liquid crystal media},}\
  }\href@noop {} {\bibfield  {journal} {\bibinfo  {journal} {Phys. Rev. Lett.}\
  }\textbf {\bibinfo {volume} {117}},\ \bibinfo {pages} {053903} (\bibinfo
  {year} {2016})}\BibitemShut {NoStop}%
\bibitem [{\citenamefont {Bliokh}\ \emph {et~al.}(2007)\citenamefont {Bliokh},
  \citenamefont {Frolov},\ and\ \citenamefont {Kravtsov}}]{Bliokh:2007}%
  \BibitemOpen
  \bibfield  {author} {\bibinfo {author} {\bibfnamefont {K.~Yu.}\ \bibnamefont
  {Bliokh}}, \bibinfo {author} {\bibfnamefont {D.~Yu.}\ \bibnamefont {Frolov}},
  \ and\ \bibinfo {author} {\bibfnamefont {Yu.~A.}\ \bibnamefont {Kravtsov}},\
  }\bibfield  {title} {\enquote {\bibinfo {title} {Non-abelian evolution of
  electromagnetic waves in a weakly anisotropic inhomogeneous medium},}\ }\href
  {\doibase 10.1103/PhysRevA.75.053821} {\bibfield  {journal} {\bibinfo
  {journal} {Phys. Rev. A}\ }\textbf {\bibinfo {volume} {75}},\ \bibinfo
  {pages} {053821} (\bibinfo {year} {2007})}\BibitemShut {NoStop}%
\bibitem [{\citenamefont {Bliokh}\ \emph {et~al.}(2008)\citenamefont {Bliokh},
  \citenamefont {Gorodetski}, \citenamefont {Kleiner},\ and\ \citenamefont
  {Hasman}}]{Bliokh:2008}%
  \BibitemOpen
  \bibfield  {author} {\bibinfo {author} {\bibfnamefont {Konstantin~Y.}\
  \bibnamefont {Bliokh}}, \bibinfo {author} {\bibfnamefont {Yuri}\ \bibnamefont
  {Gorodetski}}, \bibinfo {author} {\bibfnamefont {Vladimir}\ \bibnamefont
  {Kleiner}}, \ and\ \bibinfo {author} {\bibfnamefont {Erez}\ \bibnamefont
  {Hasman}},\ }\bibfield  {title} {\enquote {\bibinfo {title} {Coriolis effect
  in optics: Unified geometric phase and spin-{H}all effect},}\ }\href
  {\doibase 10.1103/PhysRevLett.101.030404} {\bibfield  {journal} {\bibinfo
  {journal} {Phys. Rev. Lett.}\ }\textbf {\bibinfo {volume} {101}},\ \bibinfo
  {pages} {030404} (\bibinfo {year} {2008})}\BibitemShut {NoStop}%
\bibitem [{\citenamefont {Calvo}\ and\ \citenamefont
  {Pic\'{o}n}(2007)}]{Calvo:2007}%
  \BibitemOpen
  \bibfield  {author} {\bibinfo {author} {\bibfnamefont {Gabriel~F.}\
  \bibnamefont {Calvo}}\ and\ \bibinfo {author} {\bibfnamefont {Antonio}\
  \bibnamefont {Pic\'{o}n}},\ }\bibfield  {title} {\enquote {\bibinfo {title}
  {Spin-induced angular momentum switching},}\ }\href {\doibase
  10.1364/OL.32.000838} {\bibfield  {journal} {\bibinfo  {journal} {Opt.
  Lett.}\ }\textbf {\bibinfo {volume} {32}},\ \bibinfo {pages} {838--840}
  (\bibinfo {year} {2007})}\BibitemShut {NoStop}%
\bibitem [{\citenamefont {Karimi}\ \emph {et~al.}(2009)\citenamefont {Karimi},
  \citenamefont {Piccirillo}, \citenamefont {Marrucci},\ and\ \citenamefont
  {Santamato}}]{Karimi:2009}%
  \BibitemOpen
  \bibfield  {author} {\bibinfo {author} {\bibfnamefont {Ebrahim}\ \bibnamefont
  {Karimi}}, \bibinfo {author} {\bibfnamefont {Bruno}\ \bibnamefont
  {Piccirillo}}, \bibinfo {author} {\bibfnamefont {Lorenzo}\ \bibnamefont
  {Marrucci}}, \ and\ \bibinfo {author} {\bibfnamefont {Enrico}\ \bibnamefont
  {Santamato}},\ }\bibfield  {title} {\enquote {\bibinfo {title} {Light
  propagation in a birefringent plate with topological charge},}\ }\href
  {\doibase 10.1364/OL.34.001225} {\bibfield  {journal} {\bibinfo  {journal}
  {Opt. Lett.}\ }\textbf {\bibinfo {volume} {34}},\ \bibinfo {pages}
  {1225--1227} (\bibinfo {year} {2009})}\BibitemShut {NoStop}%
\bibitem [{\citenamefont {Lax}\ \emph {et~al.}(1975)\citenamefont {Lax},
  \citenamefont {Louisell},\ and\ \citenamefont {McKnight}}]{Lax:1975}%
  \BibitemOpen
  \bibfield  {author} {\bibinfo {author} {\bibfnamefont {Melvin}\ \bibnamefont
  {Lax}}, \bibinfo {author} {\bibfnamefont {William~H.}\ \bibnamefont
  {Louisell}}, \ and\ \bibinfo {author} {\bibfnamefont {William~B.}\
  \bibnamefont {McKnight}},\ }\bibfield  {title} {\enquote {\bibinfo {title}
  {From {M}axwell to paraxial wave optics},}\ }\href@noop {} {\bibfield
  {journal} {\bibinfo  {journal} {Phys. Rev. A}\ }\textbf {\bibinfo {volume}
  {11}},\ \bibinfo {pages} {1365--1370} (\bibinfo {year} {1975})}\BibitemShut
  {NoStop}%
\bibitem [{\citenamefont {Alberucci}\ \emph {et~al.}(2016)\citenamefont
  {Alberucci}, \citenamefont {Jisha}, \citenamefont {Marrucci},\ and\
  \citenamefont {Assanto}}]{Alberucci:2016}%
  \BibitemOpen
  \bibfield  {author} {\bibinfo {author} {\bibfnamefont {A.}~\bibnamefont
  {Alberucci}}, \bibinfo {author} {\bibfnamefont {C.~P.}\ \bibnamefont
  {Jisha}}, \bibinfo {author} {\bibfnamefont {L.}~\bibnamefont {Marrucci}}, \
  and\ \bibinfo {author} {\bibfnamefont {G.}~\bibnamefont {Assanto}},\
  }\bibfield  {title} {\enquote {\bibinfo {title} {Electromagnetic confinement
  via spin-orbit interaction in anisotropic dielectrics},}\ }\href@noop {}
  {\bibfield  {journal} {\bibinfo  {journal} {ACS Photonics}\ }\textbf
  {\bibinfo {volume} {3}},\ \bibinfo {pages} {2249--2254} (\bibinfo {year}
  {2016})}\BibitemShut {NoStop}%
\bibitem [{\citenamefont {Kapitza}(1951)}]{Kapitza:1951}%
  \BibitemOpen
  \bibfield  {author} {\bibinfo {author} {\bibfnamefont {P.~L.}\ \bibnamefont
  {Kapitza}},\ }\bibfield  {title} {\enquote {\bibinfo {title} {Dynamic
  stability of a pendulum when its point of suspension vibrates},}\ }\href@noop
  {} {\bibfield  {journal} {\bibinfo  {journal} {Sov. Phys. JETP}\ }\textbf
  {\bibinfo {volume} {21}},\ \bibinfo {pages} {588--592} (\bibinfo {year}
  {1951})}\BibitemShut {NoStop}%
\bibitem [{\citenamefont {Alberucci}\ \emph {et~al.}(2013)\citenamefont
  {Alberucci}, \citenamefont {Marrucci},\ and\ \citenamefont
  {Assanto}}]{Alberucci:2013}%
  \BibitemOpen
  \bibfield  {author} {\bibinfo {author} {\bibfnamefont {A.}~\bibnamefont
  {Alberucci}}, \bibinfo {author} {\bibfnamefont {L.}~\bibnamefont {Marrucci}},
  \ and\ \bibinfo {author} {\bibfnamefont {G.}~\bibnamefont {Assanto}},\
  }\bibfield  {title} {\enquote {\bibinfo {title} {Light confinement via
  periodic modulation of the refractive index},}\ }\href
  {http://stacks.iop.org/1367-2630/15/i=8/a=083013} {\bibfield  {journal}
  {\bibinfo  {journal} {New J. Phys.}\ }\textbf {\bibinfo {volume} {15}},\
  \bibinfo {pages} {083013} (\bibinfo {year} {2013})}\BibitemShut {NoStop}%
\bibitem [{Note1()}]{Note1}%
  \BibitemOpen
  \bibinfo {note} {\protect \textit {Mutatis mutandis}, all the following
  results remain valid in the more general case of a biaxial
  crystal.}\BibitemShut {Stop}%
\bibitem [{\citenamefont {Slussarenko}\ \emph {et~al.}(2016)\citenamefont
  {Slussarenko}, \citenamefont {Alberucci}, \citenamefont {Jisha},
  \citenamefont {Piccirillo}, \citenamefont {Santamato}, \citenamefont
  {Assanto},\ and\ \citenamefont {Marrucci}}]{Slussarenko:2016}%
  \BibitemOpen
  \bibfield  {author} {\bibinfo {author} {\bibfnamefont {S.}~\bibnamefont
  {Slussarenko}}, \bibinfo {author} {\bibfnamefont {A.}~\bibnamefont
  {Alberucci}}, \bibinfo {author} {\bibfnamefont {C.~P.}\ \bibnamefont
  {Jisha}}, \bibinfo {author} {\bibfnamefont {B.}~\bibnamefont {Piccirillo}},
  \bibinfo {author} {\bibfnamefont {E.}~\bibnamefont {Santamato}}, \bibinfo
  {author} {\bibfnamefont {G.}~\bibnamefont {Assanto}}, \ and\ \bibinfo
  {author} {\bibfnamefont {L.}~\bibnamefont {Marrucci}},\ }\bibfield  {title}
  {\enquote {\bibinfo {title} {Guiding light via geometric phases},}\ }\href
  {\doibase 10.1038/NPHOTON.2016.138} {\bibfield  {journal} {\bibinfo
  {journal} {Nat. Photon.}\ } (\bibinfo {year} {2016}),\
  10.1038/NPHOTON.2016.138}\BibitemShut {NoStop}%
\bibitem [{\citenamefont {Jones}(1941)}]{Jones:1941}%
  \BibitemOpen
  \bibfield  {author} {\bibinfo {author} {\bibfnamefont {R.~Clark}\
  \bibnamefont {Jones}},\ }\bibfield  {title} {\enquote {\bibinfo {title} {A
  new calculus for the treatment of optical systemsi. description and
  discussion of the calculus},}\ }\href {\doibase 10.1364/JOSA.31.000488}
  {\bibfield  {journal} {\bibinfo  {journal} {J. Opt. Soc. Am.}\ }\textbf
  {\bibinfo {volume} {31}},\ \bibinfo {pages} {488--493} (\bibinfo {year}
  {1941})}\BibitemShut {NoStop}%
\bibitem [{Note2()}]{Note2}%
  \BibitemOpen
  \bibinfo {note} {In this Paper we adopt the the source's point of
  view.}\BibitemShut {Stop}%
\bibitem [{\citenamefont {Kwasny}\ \emph {et~al.}(2012)\citenamefont {Kwasny},
  \citenamefont {Laudyn}, \citenamefont {Sala}, \citenamefont {Alberucci},
  \citenamefont {Karpierz},\ and\ \citenamefont {Assanto}}]{Kwasny:2012}%
  \BibitemOpen
  \bibfield  {author} {\bibinfo {author} {\bibfnamefont {M.}~\bibnamefont
  {Kwasny}}, \bibinfo {author} {\bibfnamefont {U.~A.}\ \bibnamefont {Laudyn}},
  \bibinfo {author} {\bibfnamefont {F.~A.}\ \bibnamefont {Sala}}, \bibinfo
  {author} {\bibfnamefont {A.}~\bibnamefont {Alberucci}}, \bibinfo {author}
  {\bibfnamefont {M.~A.}\ \bibnamefont {Karpierz}}, \ and\ \bibinfo {author}
  {\bibfnamefont {G.}~\bibnamefont {Assanto}},\ }\bibfield  {title} {\enquote
  {\bibinfo {title} {Self-guided beams in low-birefringence nematic liquid
  crystals},}\ }\href@noop {} {\bibfield  {journal} {\bibinfo  {journal} {Phys.
  Rev. A}\ }\textbf {\bibinfo {volume} {86}},\ \bibinfo {pages} {013824}
  (\bibinfo {year} {2012})}\BibitemShut {NoStop}%
\bibitem [{\citenamefont {Dirac}(1999)}]{Dirac:1999}%
  \BibitemOpen
  \bibfield  {author} {\bibinfo {author} {\bibfnamefont {P.~A.~M.}\
  \bibnamefont {Dirac}},\ }\href@noop {} {\emph {\bibinfo {title} {The
  Principles of Quantum Mechanics}}}\ (\bibinfo  {publisher} {Oxford Science
  Publications},\ \bibinfo {address} {Oxford},\ \bibinfo {year}
  {1999})\BibitemShut {NoStop}%
\bibitem [{\citenamefont {Bliokh}\ \emph {et~al.}(2015)\citenamefont {Bliokh},
  \citenamefont {Rodr\'{i}guez-Fortuno}, \citenamefont {Nori},\ and\
  \citenamefont {Zayats}}]{Bliokh:2015}%
  \BibitemOpen
  \bibfield  {author} {\bibinfo {author} {\bibfnamefont {K.~Y.}\ \bibnamefont
  {Bliokh}}, \bibinfo {author} {\bibfnamefont {F.~J.}\ \bibnamefont
  {Rodr\'{i}guez-Fortuno}}, \bibinfo {author} {\bibfnamefont {F.}~\bibnamefont
  {Nori}}, \ and\ \bibinfo {author} {\bibfnamefont {A.~V.}\ \bibnamefont
  {Zayats}},\ }\bibfield  {title} {\enquote {\bibinfo {title} {Spin-orbit
  interactions of light},}\ }\href@noop {} {\bibfield  {journal} {\bibinfo
  {journal} {Nat. Photon.}\ }\textbf {\bibinfo {volume} {9}},\ \bibinfo {pages}
  {796--808} (\bibinfo {year} {2015})}\BibitemShut {NoStop}%
\bibitem [{Note3()}]{Note3}%
  \BibitemOpen
  \bibinfo {note} {From the standard analogy between 2D quantum mechanics and
  paraxial optics in the monochromatic regime, the propagation coordinate
  $z$plays the role of time.}\BibitemShut {Stop}%
\bibitem [{\citenamefont {Ballantine}\ \emph {et~al.}(2016)\citenamefont
  {Ballantine}, \citenamefont {Donegan},\ and\ \citenamefont
  {Eastham}}]{Ballantine:2016}%
  \BibitemOpen
  \bibfield  {author} {\bibinfo {author} {\bibfnamefont {Kyle~E.}\ \bibnamefont
  {Ballantine}}, \bibinfo {author} {\bibfnamefont {John~F.}\ \bibnamefont
  {Donegan}}, \ and\ \bibinfo {author} {\bibfnamefont {Paul~R.}\ \bibnamefont
  {Eastham}},\ }\bibfield  {title} {\enquote {\bibinfo {title} {There are many
  ways to spin a photon: Half-quantization of a total optical angular
  momentum},}\ }\href {\doibase 10.1126/sciadv.1501748} {\bibfield  {journal}
  {\bibinfo  {journal} {Science Advances}\ }\textbf {\bibinfo {volume} {2}}
  (\bibinfo {year} {2016}),\ 10.1126/sciadv.1501748},\ \Eprint
  {http://arxiv.org/abs/http://advances.sciencemag.org/content/2/4/e1501748.full.pdf}
  {http://advances.sciencemag.org/content/2/4/e1501748.full.pdf} \BibitemShut
  {NoStop}%
\bibitem [{\citenamefont {Shirley}(1965)}]{Shirley:1965}%
  \BibitemOpen
  \bibfield  {author} {\bibinfo {author} {\bibfnamefont {Jon~H.}\ \bibnamefont
  {Shirley}},\ }\bibfield  {title} {\enquote {\bibinfo {title} {Solution of the
  {S}chr\"odinger equation with a {H}amiltonian periodic in time},}\ }\href
  {\doibase 10.1103/PhysRev.138.B979} {\bibfield  {journal} {\bibinfo
  {journal} {Phys. Rev.}\ }\textbf {\bibinfo {volume} {138}},\ \bibinfo {pages}
  {B979--B987} (\bibinfo {year} {1965})}\BibitemShut {NoStop}%
\bibitem [{\citenamefont {Brown}(1991)}]{Brown:1991}%
  \BibitemOpen
  \bibfield  {author} {\bibinfo {author} {\bibfnamefont {Lowell~S.}\
  \bibnamefont {Brown}},\ }\bibfield  {title} {\enquote {\bibinfo {title}
  {Quantum motion in a paul trap},}\ }\href {\doibase
  10.1103/PhysRevLett.66.527} {\bibfield  {journal} {\bibinfo  {journal} {Phys.
  Rev. Lett.}\ }\textbf {\bibinfo {volume} {66}},\ \bibinfo {pages} {527--529}
  (\bibinfo {year} {1991})}\BibitemShut {NoStop}%
\bibitem [{Note4()}]{Note4}%
  \BibitemOpen
  \bibinfo {note} {Given the standard paraxial equation $2ik_0\protect
  \overline {n}\protect \frac {\partial A}{\partial z}+ \protect \frac
  {\partial ^2 A}{\partial x^2}+k_0^2 \Delta n^2 A=0$, the photonic potential
  is defined as $V =-k_0 \protect \frac {\Delta n^2}{2\protect \overline
  {n}}$.}\BibitemShut {Stop}%
\bibitem [{\citenamefont {Pancharatnam}(1956)}]{Pancharatnam:1956}%
  \BibitemOpen
  \bibfield  {author} {\bibinfo {author} {\bibfnamefont {S.}~\bibnamefont
  {Pancharatnam}},\ }\bibfield  {title} {\enquote {\bibinfo {title}
  {Generalized theory of interference, and its applications},}\ }\href
  {\doibase 10.1007/BF03046050} {\bibfield  {journal} {\bibinfo  {journal}
  {Proc. Indian Acad. Sci. A}\ }\textbf {\bibinfo {volume} {44}},\ \bibinfo
  {pages} {0370--0089} (\bibinfo {year} {1956})}\BibitemShut {NoStop}%
\bibitem [{\citenamefont {Berry}(1994)}]{Berry:1994}%
  \BibitemOpen
  \bibfield  {author} {\bibinfo {author} {\bibfnamefont {M.~V.}\ \bibnamefont
  {Berry}},\ }\bibfield  {title} {\enquote {\bibinfo {title} {{Pancharatnam},
  virtuoso of the {P}oincar\'e sphere: an appreciation},}\ }\href@noop {}
  {\bibfield  {journal} {\bibinfo  {journal} {Current Science}\ }\textbf
  {\bibinfo {volume} {67}},\ \bibinfo {pages} {220--223} (\bibinfo {year}
  {1994})}\BibitemShut {NoStop}%
\bibitem [{\citenamefont {Bliokh}\ \emph {et~al.}(2016)\citenamefont {Bliokh},
  \citenamefont {Samlan}, \citenamefont {Prajapati}, \citenamefont {Puentes},
  \citenamefont {Viswanathan},\ and\ \citenamefont {Nori}}]{Bliokh:2016}%
  \BibitemOpen
  \bibfield  {author} {\bibinfo {author} {\bibfnamefont {K.~Y.}\ \bibnamefont
  {Bliokh}}, \bibinfo {author} {\bibfnamefont {C.~T.}\ \bibnamefont {Samlan}},
  \bibinfo {author} {\bibfnamefont {C.}~\bibnamefont {Prajapati}}, \bibinfo
  {author} {\bibfnamefont {G.}~\bibnamefont {Puentes}}, \bibinfo {author}
  {\bibfnamefont {N.~K.}\ \bibnamefont {Viswanathan}}, \ and\ \bibinfo {author}
  {\bibfnamefont {F.}~\bibnamefont {Nori}},\ }\bibfield  {title} {\enquote
  {\bibinfo {title} {Spin-hall effect and circular birefringence of a uniaxial
  crystal plate},}\ }\href@noop {} {\bibfield  {journal} {\bibinfo  {journal}
  {Optica}\ }\textbf {\bibinfo {volume} {3}},\ \bibinfo {pages} {1039--1047}
  (\bibinfo {year} {2016})}\BibitemShut {NoStop}%
\bibitem [{\citenamefont {Hu}\ and\ \citenamefont {Menyuk}(2009)}]{Hu:2009}%
  \BibitemOpen
  \bibfield  {author} {\bibinfo {author} {\bibfnamefont {Jonathan}\
  \bibnamefont {Hu}}\ and\ \bibinfo {author} {\bibfnamefont {Curtis~R.}\
  \bibnamefont {Menyuk}},\ }\bibfield  {title} {\enquote {\bibinfo {title}
  {Understanding leaky modes: slab waveguide revisited},}\ }\href {\doibase
  10.1364/AOP.1.000058} {\bibfield  {journal} {\bibinfo  {journal} {Adv. Opt.
  Photon.}\ }\textbf {\bibinfo {volume} {1}},\ \bibinfo {pages} {58--106}
  (\bibinfo {year} {2009})}\BibitemShut {NoStop}%
\bibitem [{\citenamefont {Oskooi}\ \emph {et~al.}(2010)\citenamefont {Oskooi},
  \citenamefont {Roundy}, \citenamefont {Ibanescu}, \citenamefont {Bermel},
  \citenamefont {Joannopoulos},\ and\ \citenamefont {Johnson}}]{Oskooi:2010}%
  \BibitemOpen
  \bibfield  {author} {\bibinfo {author} {\bibfnamefont {Ardavan~F.}\
  \bibnamefont {Oskooi}}, \bibinfo {author} {\bibfnamefont {David}\
  \bibnamefont {Roundy}}, \bibinfo {author} {\bibfnamefont {Mihai}\
  \bibnamefont {Ibanescu}}, \bibinfo {author} {\bibfnamefont {Peter}\
  \bibnamefont {Bermel}}, \bibinfo {author} {\bibfnamefont {J.~D.}\
  \bibnamefont {Joannopoulos}}, \ and\ \bibinfo {author} {\bibfnamefont
  {Steven~G.}\ \bibnamefont {Johnson}},\ }\bibfield  {title} {\enquote
  {\bibinfo {title} {{MEEP}: A flexible free-software package for
  electromagnetic simulations by the {FDTD} method},}\ }\href {\doibase
  doi:10.1016/j.cpc.2009.11.008} {\bibfield  {journal} {\bibinfo  {journal}
  {Comput. Phys. Commun.}\ }\textbf {\bibinfo {volume} {181}},\ \bibinfo
  {pages} {687--702} (\bibinfo {year} {2010})}\BibitemShut {NoStop}%
\bibitem [{\citenamefont {D'Alessandro}\ \emph {et~al.}(2015)\citenamefont
  {D'Alessandro}, \citenamefont {Martini}, \citenamefont {Gilardi},
  \citenamefont {Beccherelli},\ and\ \citenamefont
  {Asquini}}]{Dalessandro:2015}%
  \BibitemOpen
  \bibfield  {author} {\bibinfo {author} {\bibfnamefont {A.}~\bibnamefont
  {D'Alessandro}}, \bibinfo {author} {\bibfnamefont {L.}~\bibnamefont
  {Martini}}, \bibinfo {author} {\bibfnamefont {G.}~\bibnamefont {Gilardi}},
  \bibinfo {author} {\bibfnamefont {R.}~\bibnamefont {Beccherelli}}, \ and\
  \bibinfo {author} {\bibfnamefont {R.}~\bibnamefont {Asquini}},\ }\bibfield
  {title} {\enquote {\bibinfo {title} {Polarization-independent nematic liquid
  crystal waveguides for optofluidic applications},}\ }\href {\doibase
  10.1109/LPT.2015.2438151} {\bibfield  {journal} {\bibinfo  {journal} {IEEE
  Photon. Techn. Lett.}\ }\textbf {\bibinfo {volume} {27}},\ \bibinfo {pages}
  {1709--1712} (\bibinfo {year} {2015})}\BibitemShut {NoStop}%
\bibitem [{\citenamefont {Askarpour}\ \emph {et~al.}(2014)\citenamefont
  {Askarpour}, \citenamefont {Zhao},\ and\ \citenamefont
  {Al\`u}}]{Askarpour:2014}%
  \BibitemOpen
  \bibfield  {author} {\bibinfo {author} {\bibfnamefont {Amir~Nader}\
  \bibnamefont {Askarpour}}, \bibinfo {author} {\bibfnamefont {Yang}\
  \bibnamefont {Zhao}}, \ and\ \bibinfo {author} {\bibfnamefont {Andrea}\
  \bibnamefont {Al\`u}},\ }\bibfield  {title} {\enquote {\bibinfo {title} {Wave
  propagation in twisted metamaterials},}\ }\href {\doibase
  10.1103/PhysRevB.90.054305} {\bibfield  {journal} {\bibinfo  {journal} {Phys.
  Rev. B}\ }\textbf {\bibinfo {volume} {90}},\ \bibinfo {pages} {054305}
  (\bibinfo {year} {2014})}\BibitemShut {NoStop}%
\bibitem [{\citenamefont {Pendry}\ \emph {et~al.}(1999)\citenamefont {Pendry},
  \citenamefont {Holden}, \citenamefont {Robbins},\ and\ \citenamefont
  {Stewart}}]{Pendry:1999}%
  \BibitemOpen
  \bibfield  {author} {\bibinfo {author} {\bibfnamefont {J.B.}\ \bibnamefont
  {Pendry}}, \bibinfo {author} {\bibfnamefont {A.J.}\ \bibnamefont {Holden}},
  \bibinfo {author} {\bibfnamefont {D.J.}\ \bibnamefont {Robbins}}, \ and\
  \bibinfo {author} {\bibfnamefont {W.J.}\ \bibnamefont {Stewart}},\ }\bibfield
   {title} {\enquote {\bibinfo {title} {Magnetism from conductors and enhanced
  nonlinear phenomena},}\ }\href {\doibase 10.1109/22.798002} {\bibfield
  {journal} {\bibinfo  {journal} {Microwave Theory and Techniques, IEEE
  Transactions on}\ }\textbf {\bibinfo {volume} {47}},\ \bibinfo {pages}
  {2075--2084} (\bibinfo {year} {1999})}\BibitemShut {NoStop}%
\bibitem [{\citenamefont {Smith}\ and\ \citenamefont
  {Schurig}(2003)}]{Smith:2003}%
  \BibitemOpen
  \bibfield  {author} {\bibinfo {author} {\bibfnamefont {D.~R.}\ \bibnamefont
  {Smith}}\ and\ \bibinfo {author} {\bibfnamefont {D.}~\bibnamefont
  {Schurig}},\ }\bibfield  {title} {\enquote {\bibinfo {title} {Electromagnetic
  wave propagation in media with indefinite permittivity and permeability
  tensors},}\ }\href {\doibase 10.1103/PhysRevLett.90.077405} {\bibfield
  {journal} {\bibinfo  {journal} {Phys. Rev. Lett.}\ }\textbf {\bibinfo
  {volume} {90}},\ \bibinfo {pages} {077405} (\bibinfo {year}
  {2003})}\BibitemShut {NoStop}%
\bibitem [{\citenamefont {Ghosh}\ \emph {et~al.}(2008)\citenamefont {Ghosh},
  \citenamefont {Sheridon},\ and\ \citenamefont {Fischer}}]{Ghosh:2008}%
  \BibitemOpen
  \bibfield  {author} {\bibinfo {author} {\bibfnamefont {A.}~\bibnamefont
  {Ghosh}}, \bibinfo {author} {\bibfnamefont {N.~K.}\ \bibnamefont {Sheridon}},
  \ and\ \bibinfo {author} {\bibfnamefont {P.}~\bibnamefont {Fischer}},\
  }\bibfield  {title} {\enquote {\bibinfo {title} {Voltage-controllable
  magnetic composite based on multifunctional polyethylene microparticles},}\
  }\href {\doibase 10.1002/smll.200701301} {\bibfield  {journal} {\bibinfo
  {journal} {Small}\ }\textbf {\bibinfo {volume} {4}},\ \bibinfo {pages}
  {1956--1958} (\bibinfo {year} {2008})}\BibitemShut {NoStop}%
\bibitem [{\citenamefont {Khanikaev}\ \emph {et~al.}(2013)\citenamefont
  {Khanikaev}, \citenamefont {Mousavi}, \citenamefont {Tse}, \citenamefont
  {Kargarian}, \citenamefont {MacDonald},\ and\ \citenamefont
  {Shvets}}]{Khanikaev:2013}%
  \BibitemOpen
  \bibfield  {author} {\bibinfo {author} {\bibfnamefont {Alexander~B.}\
  \bibnamefont {Khanikaev}}, \bibinfo {author} {\bibfnamefont {S.~Hossein}\
  \bibnamefont {Mousavi}}, \bibinfo {author} {\bibfnamefont {Wang-Kong}\
  \bibnamefont {Tse}}, \bibinfo {author} {\bibfnamefont {Mehdi}\ \bibnamefont
  {Kargarian}}, \bibinfo {author} {\bibfnamefont {Allan~H.}\ \bibnamefont
  {MacDonald}}, \ and\ \bibinfo {author} {\bibfnamefont {Gennady}\ \bibnamefont
  {Shvets}},\ }\bibfield  {title} {\enquote {\bibinfo {title} {Photonic
  topological insulator},}\ }\href@noop {} {\bibfield  {journal} {\bibinfo
  {journal} {Nat. Mater.}\ }\textbf {\bibinfo {volume} {12}},\ \bibinfo {pages}
  {233--239} (\bibinfo {year} {2013})}\BibitemShut {NoStop}%
\bibitem [{\citenamefont {Pfeiffer}\ and\ \citenamefont
  {Grbic}(2013)}]{Pfeiffer:2013}%
  \BibitemOpen
  \bibfield  {author} {\bibinfo {author} {\bibfnamefont {Carl}\ \bibnamefont
  {Pfeiffer}}\ and\ \bibinfo {author} {\bibfnamefont {Anthony}\ \bibnamefont
  {Grbic}},\ }\bibfield  {title} {\enquote {\bibinfo {title} {Metamaterial
  {H}uygens' surfaces: Tailoring wave fronts with reflectionless sheets},}\
  }\href {\doibase 10.1103/PhysRevLett.110.197401} {\bibfield  {journal}
  {\bibinfo  {journal} {Phys. Rev. Lett.}\ }\textbf {\bibinfo {volume} {110}},\
  \bibinfo {pages} {197401} (\bibinfo {year} {2013})}\BibitemShut {NoStop}%
\bibitem [{\citenamefont {Bliokh}\ \emph {et~al.}(2014)\citenamefont {Bliokh},
  \citenamefont {Kivshar},\ and\ \citenamefont {Nori}}]{Bliokh:2014}%
  \BibitemOpen
  \bibfield  {author} {\bibinfo {author} {\bibfnamefont {Konstantin~Y.}\
  \bibnamefont {Bliokh}}, \bibinfo {author} {\bibfnamefont {Yuri~S.}\
  \bibnamefont {Kivshar}}, \ and\ \bibinfo {author} {\bibfnamefont {Franco}\
  \bibnamefont {Nori}},\ }\bibfield  {title} {\enquote {\bibinfo {title}
  {Magnetoelectric effects in local light-matter interactions},}\ }\href
  {\doibase 10.1103/PhysRevLett.113.033601} {\bibfield  {journal} {\bibinfo
  {journal} {Phys. Rev. Lett.}\ }\textbf {\bibinfo {volume} {113}},\ \bibinfo
  {pages} {033601} (\bibinfo {year} {2014})}\BibitemShut {NoStop}%
\bibitem [{\citenamefont {Asadchy}\ \emph {et~al.}(2015)\citenamefont
  {Asadchy}, \citenamefont {Ra'di}, \citenamefont {Vehmas},\ and\ \citenamefont
  {Tretyakov}}]{Asadchy:2015}%
  \BibitemOpen
  \bibfield  {author} {\bibinfo {author} {\bibfnamefont {V.~S.}\ \bibnamefont
  {Asadchy}}, \bibinfo {author} {\bibfnamefont {Y.}~\bibnamefont {Ra'di}},
  \bibinfo {author} {\bibfnamefont {J.}~\bibnamefont {Vehmas}}, \ and\ \bibinfo
  {author} {\bibfnamefont {S.~A.}\ \bibnamefont {Tretyakov}},\ }\bibfield
  {title} {\enquote {\bibinfo {title} {Functional metamirrors using
  bianisotropic elements},}\ }\href {\doibase 10.1103/PhysRevLett.114.095503}
  {\bibfield  {journal} {\bibinfo  {journal} {Phys. Rev. Lett.}\ }\textbf
  {\bibinfo {volume} {114}},\ \bibinfo {pages} {095503} (\bibinfo {year}
  {2015})}\BibitemShut {NoStop}%
\bibitem [{\citenamefont {Fang}\ and\ \citenamefont {Fan}(2013)}]{Fang:2013}%
  \BibitemOpen
  \bibfield  {author} {\bibinfo {author} {\bibfnamefont {Kejie}\ \bibnamefont
  {Fang}}\ and\ \bibinfo {author} {\bibfnamefont {Shanhui}\ \bibnamefont
  {Fan}},\ }\bibfield  {title} {\enquote {\bibinfo {title} {Controlling the
  flow of light using the inhomogeneous effective gauge field that emerges from
  dynamic modulation},}\ }\href {\doibase 10.1103/PhysRevLett.111.203901}
  {\bibfield  {journal} {\bibinfo  {journal} {Phys. Rev. Lett.}\ }\textbf
  {\bibinfo {volume} {111}},\ \bibinfo {pages} {203901} (\bibinfo {year}
  {2013})}\BibitemShut {NoStop}%
\bibitem [{\citenamefont {Lin}\ and\ \citenamefont {Fan}(2014)}]{Lin:2014_1}%
  \BibitemOpen
  \bibfield  {author} {\bibinfo {author} {\bibfnamefont {Qian}\ \bibnamefont
  {Lin}}\ and\ \bibinfo {author} {\bibfnamefont {Shanhui}\ \bibnamefont
  {Fan}},\ }\bibfield  {title} {\enquote {\bibinfo {title} {Light guiding by
  effective gauge field for photons},}\ }\href {\doibase
  10.1103/PhysRevX.4.031031} {\bibfield  {journal} {\bibinfo  {journal} {Phys.
  Rev. X}\ }\textbf {\bibinfo {volume} {4}},\ \bibinfo {pages} {031031}
  (\bibinfo {year} {2014})}\BibitemShut {NoStop}%
\bibitem [{\citenamefont {Paul}(1990)}]{Paul:1990}%
  \BibitemOpen
  \bibfield  {author} {\bibinfo {author} {\bibfnamefont {Wolfgang}\
  \bibnamefont {Paul}},\ }\bibfield  {title} {\enquote {\bibinfo {title}
  {Electromagnetic traps for charged and neutral particles},}\ }\href {\doibase
  10.1103/RevModPhys.62.531} {\bibfield  {journal} {\bibinfo  {journal} {Rev.
  Mod. Phys.}\ }\textbf {\bibinfo {volume} {62}},\ \bibinfo {pages} {531--540}
  (\bibinfo {year} {1990})}\BibitemShut {NoStop}%
\end{thebibliography}

%

\end{document}